# Granger Mediation Analysis of Multiple Time Series with an Application to fMRI

**Yi Zhao**

Department of Biostatistics, Brown University, Providence, Rhode Island, U.S.A.

*email:* yi_zhao@alumni.brown.edu

**and**

**Xi Luo**

Department of Biostatistics, Brown University, Providence, Rhode Island, U.S.A.

*email:* xi.rossi.luo@gmail.com

SUMMARY: It becomes increasingly popular to perform mediation analysis for complex data from sophisticated experimental studies. In this paper, we present Granger Mediation Analysis (GMA), a new framework for causal mediation analysis of multiple time series. This framework is motivated by a functional magnetic resonance imaging (fMRI) experiment where we are interested in estimating the mediation effects between a randomized stimulus time series and brain activity time series from two brain regions. The stable unit treatment assumption for causal mediation analysis is thus unrealistic for this type of time series data. To address this challenge, our framework integrates two types of models: causal mediation analysis across the variables and vector autoregressive models across the temporal observations. We further extend this framework to handle multilevel data to address individual variability and correlated errors between the mediator and the outcome variables. These models not only provide valid causal mediation for time series data but also model the causal dynamics across time. We show that the modeling parameters in our models are identifiable, and we develop computationally efficient methods to maximize the likelihood-based optimization criteria. Simulation studies show that our method reduces the estimation bias and improve statistical power, compared to existing approaches. On a real fMRI data set, our approach not only infers the causal effects of brain pathways but accurately captures the feedback effect of the outcome region on the mediator region.

This paper has been submitted for consideration for publication in *Biometrics*

## 1. Introduction

Mediation analysis is a popular statistical approach for many social and scientific studies. It aims to assess the role of an intermediate variable or mediator sitting in the pathway from a treatment variable to an outcome variable. In many studies, observations from multiple units or subjects are collected, and existing mediation methods usually impose the assumption of independent units explicitly or implicitly. For example, the original Baron-Kenny method (Baron and Kenny, 1986; MacKinnon, 2008), under the structural equation modeling framework, relies on the independence assumption to carry out estimation and inference. Causal mediation analysis, widely studied in the statistical literature, was developed to infer the causal effects in mediation models, see a review Imai et al. (2010). Under the potential outcomes framework for causal inference (Rubin, 1974), most methods require additional assumptions. One assumption, the Stable Unit Treatment Value Assumption (SUTVA) (Rubin, 1978, 1980), implies the independence of units or "no interference". All these methods mentioned before, however, cannot be applied to time series data, because clearly the independence assumption is violated.

In this paper, we will focus on the time series data generated from a functional Magnetic Resonance Imaging (fMRI) experiment when each participant performs a motor task responding to randomized experimental stimuli. During the experiment, brain activities are measured by fMRI using the blood-oxygen level dependent (BOLD) contrast. Here, we are interested in quantifying how the activities in the presupplementary motor area (preSMA) mediate the activities in the primary motor cortex (M1) responding to the stimulus input series following a standard BOLD model (Friston, 2009). In this data example, the stimulus is the treatment variable, and the BOLD activities in preSMA and M1 are the mediator and outcome variables respectively. All these variables are time series from each participant, and an example of these three time series from one participant is shown in Figure 1.

[Figure 1 about here.]

It has been well established before that BOLD time series by fMRI have non-ignorable temporal correlations, which can be effectively modeled by autoregressive (AR) models with a small lag order, see the review Lindquist (2008).

Indeed, autoregressive modeling is an important approach for time series analysis, especially for fMRI data. One earlier approach, named as Granger causality (Granger, 1969, 1980), assesses if the current value of time series $x$ can be predicted by the past values of time series $x$ and another time series $y$. This approach is later generalized for multiple time series by multivariate autoregressive models (Harrison et al., 2003; Goebel et al., 2003). All these methods use prediction to assess the temporal relationships between multiple variables, without using the potential outcomes framework to define causality formally. More importantly, they cannot quantify the pathway effects as in mediation analysis.

Mediation analysis for fMRI data is becoming an increasingly popular topic. Atlas et al. (2010) applied mediation analysis to study the brain mediators of a self-reported behavioral outcome. They utilized a general linear model (GLM) approach to model the coefficients for brain activities or single-trial betas, and thus these coefficients in their mediation model may be considered independent assuming that the temporal correlations in fMRI time series are removed by the GLM. Lindquist (2012) proposed a functional mediation model with fMRI mediator and a scalar and non-time series outcome. With also a scalar behavioral outcome, Chén et al. (2015) recently proposed multiple mediator models where none of the mediators is modeled as time series. Using the single-trial beta approach, Zhao and Luo (2014) proposed a multilevel causal mediation framework for single-trial betas as the mediator and outcome. It addresses the unmeasured confounding and individual variation issues, but did not directly model the temporal dependence in fMRI time series as all the other fMRI mediation methods mentioned before.

In a related data setting for longitudinal data, marginal structural models for causal mediation analysis were proposed to infer causal pathways for time-varying treatments and mediators (Robins et al., 2000; van der Laan and Petersen, 2008; VanderWeele, 2009). These marginal models usually study a single outcome at one time point, which is also different from our time series setting. They were also developed mostly for longitudinal data with fewer temporal observations, and thus the temporal dependence relationships are not the focus of modeling or experiments.

We address these methodological limitations by proposing a new framework, called Granger Mediation Analysis (GMA). It allows modeling time series for all the three variables in mediation analysis. A conceptual diagram of our model is illustrated in Figure 2.

[Figure 2 about here.]

At each time point $t$, randomized treatment $Z_t$ influences both mediator $M_t$ and outcome $R_t$ as a mediation model, and the errors in the model are autoregressive across time. In our fMRI example, $Z_t$ is the randomized stimulus input, $M_t$ and $R_t$ are the BOLD time series from the preSMA and M1. To some extent, our model thus aims to infer the temporal and spatial dependence in fMRI time series, which is an important scientific problem (Roebroeck et al., 2005; Eichler, 2005; Londei et al., 2006; Friston, 2009). We further develop the multilevel modeling idea in Zhao and Luo (2014) to model individual variations across subjects and to correct for the bias introduced by unmeasured confounders $U_t$. This bias is an important issue for fMRI mediation analysis, as many papers have shown that there exist unmeasured or unmodeled factors in fMRI experiments that influence BOLD signals across the brain, see Fox et al. (2006), Fair et al. (2007), Mason et al. (2007) and the discussion in Zhao and Luo (2014).

This paper is organized as follows. In Section 2, we introduce the Granger Mediation Analysis framework, which consists of a lower-level mediation model for individual time series (Section 2.1) and a regression model for population-level causal inference (Section 2.4). We compare our method with existing methods through simulation studies in Section 3 and an analysis of the fMRI data set in Section 4. Section 5 summarizes this paper with discussions and future work.

## 2. Model

### 2.1 *A mediation model for time series*

In this section, we first introduce our single-level GMA model for time series data, and this is our model for the fMRI time series for each participant $i$. We will extend this model to multi-level data from multiple participants in Section 2.4. To keep the following discussion uncluttered, we drop the participant index $i$ hereafter for this first level model.

Using the structural equation modeling (SEM) framework, our mediation model is first written as

$$M_t = Z_t A + E_{1t}, \quad (1)$$

$$R_t = Z_t C + M_t B + E_{2t}, \quad (2)$$

where $Z_t$ is the random treatment assignment at time $t$, $R_t$ and $M_t$ are the outcomes; $A$, $C$ and $B$ are the coefficients of interest; and $E_{1t}$ and $E_{2t}$ are the model errors. To characterize the interregional and temporal dependencies, we propose a multivariate autoregressive model of order $p$ (MAR($p$)) for $(E_{1t}, E_{2t})$ as

$$\begin{aligned} E_{1t} &= \sum_{j=1}^{p} \omega_{11_j} E_{1,t-j} + \sum_{j=1}^{p} \omega_{21_j} E_{2,t-j} + \epsilon_{1t}, \\ E_{2t} &= \sum_{j=1}^{p} \omega_{12_j} E_{1,t-j} + \sum_{j=1}^{p} \omega_{22_j} E_{2,t-j} + \epsilon_{2t}, \end{aligned} \quad (3)$$

where the error vector $(\epsilon_{1t}, \epsilon_{2t})^\top$ is assumed to be a Gaussian white noise process as

$$\begin{pmatrix} \epsilon_{1t} \\ \epsilon_{2t} \end{pmatrix} \sim \mathcal{N}(\mathbf{0}, \mathbf{\Sigma}), \quad \mathbf{\Sigma} = \begin{pmatrix} \sigma_1^2 & \delta\sigma_1\sigma_2 \\ \delta\sigma_1\sigma_2 & \sigma_2^2 \end{pmatrix},$$

and $(\epsilon_{1t}, \epsilon_{2t})^\top$ is independent of $(\epsilon_{1s}, \epsilon_{2s})^\top$ for $s \neq t$. Standard mediation models take the same form in appearance as ours in (1) and (2), but usually under the assumption that the two errors are mutually independent and independent across $t$. Here, we introduce a temporal dependence of the errors using (3), and a mediator-outcome dependence due to unmeasured confounding using $\delta \neq 0$ in $\boldsymbol{\Sigma}$. We illustrate the latter point using the following model with an unmeasured confounding factor $U_t$ as

$$\begin{aligned} E_{1t} &= \sum_{j=1}^{p} \omega_{11_j} E_{1,t-j} + \sum_{j=1}^{p} \omega_{21_j} E_{2,t-j} + g_1 U_t + \tilde{e}_{1t}, \\ E_{2t} &= \sum_{j=1}^{p} \omega_{12_j} E_{1,t-j} + \sum_{j=1}^{p} \omega_{22_j} E_{2,t-j} + g_2 U_t + \tilde{e}_{2t}, \end{aligned} \tag{4}$$

where $U_t$, $\tilde{e}_{1t}$ and $\tilde{e}_{2t}$ are mutually independent stochastic processes; $U_t$ is independent of $U_s$, and $\tilde{e}_{rt}$ is independent of $\tilde{e}_{rs}$, for $s \neq t$ and $r = 1, 2$. Under this additive confounding effect model, $E_{1t}$ and $E_{2t}$ are correlated when $g_1 g_2 \neq 0$. In fMRI, we expect $\delta \neq 0$ due to various confounding factors discussed in Zhao and Luo (2014).

We impose the following stationary condition: the eigenvalues of the companion matrix

$$\boldsymbol{F} = \begin{pmatrix} \boldsymbol{\Omega}_1^\top & \boldsymbol{\Omega}_2^\top & \cdots & \boldsymbol{\Omega}_{p-1}^\top & \boldsymbol{\Omega}_p^\top \\ \mathbf{I}_2 & \mathbf{0} & \cdots & \mathbf{0} & \mathbf{0} \\ \mathbf{0} & \mathbf{I}_2 & \cdots & \mathbf{0} & \mathbf{0} \\ \vdots & \vdots & \ddots & \vdots & \vdots \\ \mathbf{0} & \mathbf{0} & \cdots & \mathbf{I}_2 & \mathbf{0} \end{pmatrix}, \quad \text{where } \boldsymbol{\Omega}_j = \begin{pmatrix} \omega_{11_j} & \omega_{12_j} \\ \omega_{21_j} & \omega_{22_j} \end{pmatrix},$$

have modulus less than one. This is a standard condition for stationary autoregressive processes, see the textbook Shumway and Stoffer (2010) for example. This stationary condition is deemed satisfied for fMRI data after removing the stimulus effects, as in our model, see Harrison et al. (2003); Chang and Glover (2010).

### 2.2 *Assumptions and causal interpretation*

Using Rubin's potential outcome framework (Rubin, 2005), we impose the following causal assumptions:

(A1) the treatment assignment regime is the same at each time point $t$;

(A2) models (1), (2) and (3) are correctly specified;

(A3) at each time point $t$, the observed outcome is one realization of the potential outcome with observed treatment assignment $z$;

(A4) $Z_t$ is sequentially randomly assigned with $0 < \mathbb{P}(Z_t = z) < 1$ for every $z$,

$$\{R_t((z', \{z_s\}_{s<t}), \{m_t\}_t), M_t((z, \{z_s\}_{s<t})\} \perp\!\!\!\perp Z_t \mid \{Z_s = z_s\}_{s<t},$$

$$\{E_{1t}(z, \{z_s\}_{s<t}), E_{2t}(z', \{z_s\}_{s<t})\} \perp\!\!\!\perp Z_t \mid \{Z_s = z_s\}_{s<t},$$

and $\{\epsilon_{1t}(z), \epsilon_{2t}(z')\} \perp\!\!\!\perp Z_t$ for $\forall z, z'$;

(A5) the unmeasured confounding factor $U_t$ affects the outcome only at time $t$, and this impact is assumed to be additive and linear.

Assumption (A1) is expected to hold in our experiment because the treatment $Z_t$ is randomized. The stable unit treatment value assumption (SUTVA) (Rubin, 1978, 1980) actually includes (A1) and additionally that the potential outcome of one trial is unrelated to the treatment assignment of other trials. The latter is not satisfied for time series, and thus we consider this relaxed assumption (A1). Assumptions (A2)-(A4) are standard regularity assumptions in causal mediation inference (Rubin, 1978; Holland, 1988; Robins, 2000; Imai et al., 2010; Imai and Yamamoto, 2013; VanderWeele, 2015). Similar to the multilevel framework in Zhao and Luo (2014), (A5) relaxes the standard assumption on the ignorability of the mediator. In Section A.1 of the supplementary materials, we show that the causal estimands are identifiable. Assumptions (A1)-(A5), together with the stationarity assumption of $\{(E_{1t}, E_{2t})\}_t$, guarantee that our proposed estimators consistently estimate the causal effects, see Theorem 2 and Theorem A.1 of the supplementary materials.

## 2.3 *Method*

Given models (1), (2) and (3), it is difficult to derive the explicit form of the joint distribution of $(M_t, R_t)$, because of the spatio-temporal dependence between $E_{1t}$ and $E_{2t}$. We first introduce the following equivalent formulation of our GMA model

$$M_t = Z_t A + \sum_{j=1}^{p} \phi_{1j} Z_{t-j} + \sum_{j=1}^{p} \psi_{11_j} M_{t-j} + \sum_{j=1}^{p} \psi_{21_j} R_{t-j} + \epsilon_{1t}, \tag{5}$$

$$R_t = Z_t C + M_t B + \sum_{j=1}^{p} \phi_{2j} Z_{t-j} + \sum_{j=1}^{p} \psi_{12_j} M_{t-j} + \sum_{j=1}^{p} \psi_{22_j} R_{t-j} + \epsilon_{2t}, \tag{6}$$

with

$$\boldsymbol{\eta}_j \triangleq \begin{pmatrix} \phi_{1j} \\ \phi_{2j} \\ \psi_{11_j} \\ \psi_{21_j} \\ \psi_{12_j} \\ \psi_{22_j} \end{pmatrix} = \begin{pmatrix} -A & -C & 0 & 0 \\ 0 & 0 & -A & -C \\ 1 & -B & 0 & 0 \\ 0 & 1 & 0 & 0 \\ 0 & 0 & 1 & -B \\ 0 & 0 & 0 & 1 \end{pmatrix} \begin{pmatrix} \omega_{11_j} \\ \omega_{21_j} \\ \omega_{12_j} \\ \omega_{22_j} \end{pmatrix} \triangleq \boldsymbol{D}\boldsymbol{\omega}_j. \tag{7}$$

This new formulation shows that both $M_t$ and $R_t$ not only can be influenced by $Z_t$, but also by $(Z_{t-1}, M_{t-1}, R_{t-1}, \ldots, Z_{t-p}, M_{t-p}, R_{t-p})$. This allows the treatment to affect future observations in the time series setting, even if the treatment has already ended. Figure 3 illustrates the causal diagram of models (5) and (6) using the case of $p = 1$.

[Figure 3 about here.]

Besides the advantage of this interpretation, another benefit of models (5) and (6) is that $(\epsilon_{1t}, \epsilon_{2t})$ are Gaussian white noises, which are uncorrelated in time. We thus propose to maximize the conditional likelihood (see Shumway and Stoffer (2010)) to estimate the parameters, conditioning on the initial $p$ observations. This is a popular approach for time series estimation, because it eases the computation and implementation burden for general $p$ and usually yields approximately the same estimates as maximizing the unconditional likelihood.

This formulation essentially absorbs the autoregressive model (3) into models (1) and (2), and the new modeling parameter $\boldsymbol{\eta}_j$ is a linear transformation of the autoregressive model parameter $\boldsymbol{\omega}_j$ as stated in (7). Reversely, it is easy to show that $\boldsymbol{D}$ has full column rank and thus $\boldsymbol{\omega}_j$ is recovered from the parameters in the new formulation.

LEMMA 1: *Given any $\boldsymbol{\eta}_j$ and $(A, B, C)$, $\boldsymbol{\omega}_j$ is uniquely determined by $\boldsymbol{\omega}_j = (\boldsymbol{D}^T\boldsymbol{D})^{-1}\boldsymbol{D}^T\boldsymbol{\eta}_j$, and $\boldsymbol{D}^T\boldsymbol{D}$ is always invertible.*

Based on Lemma 1, we propose to estimate first the parameters in models (5) and (6), i.e., $\{A, B, C, (\boldsymbol{\eta}_j)\}$. To simplify the notation, we introduce the following matrix representations:

$$\boldsymbol{\theta}_1 = \begin{pmatrix} A \\ \boldsymbol{\phi}_1 \\ \boldsymbol{\psi}_{11} \\ \boldsymbol{\psi}_{21} \end{pmatrix}, \quad \boldsymbol{\theta}_2 = \begin{pmatrix} C \\ \boldsymbol{\phi}_2 \\ \boldsymbol{\psi}_{12} \\ \boldsymbol{\psi}_{22} \end{pmatrix}, \quad \boldsymbol{\phi}_1 = \begin{pmatrix} \phi_{1_1} \\ \vdots \\ \phi_{1_p} \end{pmatrix}, \quad \boldsymbol{\phi}_2 = \begin{pmatrix} \phi_{2_1} \\ \vdots \\ \phi_{2_p} \end{pmatrix},$$

$$\boldsymbol{\psi}_{11} = \begin{pmatrix} \psi_{11_1} \\ \vdots \\ \psi_{11_p} \end{pmatrix}, \quad \boldsymbol{\psi}_{21} = \begin{pmatrix} \psi_{21_1} \\ \vdots \\ \psi_{22_p} \end{pmatrix}, \quad \boldsymbol{\psi}_{12} = \begin{pmatrix} \psi_{12_1} \\ \vdots \\ \psi_{12_p} \end{pmatrix}, \quad \boldsymbol{\psi}_{22} = \begin{pmatrix} \psi_{22_1} \\ \vdots \\ \psi_{22_p} \end{pmatrix},$$

$$\boldsymbol{Z}_{t-1}^{(p)} = \begin{pmatrix} Z_{t-1} \\ \vdots \\ Z_{t-p} \end{pmatrix}, \quad \boldsymbol{M}_{t-1}^{(p)} = \begin{pmatrix} M_{t-1} \\ \vdots \\ M_{t-p} \end{pmatrix}, \quad \boldsymbol{R}_{t-1}^{(p)} = \begin{pmatrix} R_{t-1} \\ \vdots \\ R_{t-p} \end{pmatrix}, \quad \boldsymbol{X}_t = \begin{pmatrix} Z_t \\ \boldsymbol{Z}_{t-1}^{(p)} \\ \boldsymbol{M}_{t-1}^{(p)} \\ \boldsymbol{R}_{t-1}^{(p)} \end{pmatrix}.$$

Models (5) and (6) can be written in the following form

$$M_t = \boldsymbol{X}_t^\top \boldsymbol{\theta}_1 + \epsilon_{1t}, \tag{8}$$

$$R_t = M_t B + \boldsymbol{X}_t^\top \boldsymbol{\theta}_2 + \epsilon_{2t}. \tag{9}$$

Given the initial $p$ observations, the conditional log-likelihood (ignoring constants) is

$$\begin{aligned}\ell\left(\boldsymbol{\theta}_1, \boldsymbol{\theta}_2, B, \boldsymbol{\Sigma}\right) = & -\frac{T-p}{2}\log\sigma_1^2\sigma_2^2(1-\delta^2) - \frac{1}{2\sigma_1^2}\|\boldsymbol{M} - \boldsymbol{X}\boldsymbol{\theta}_1\|_2^2 \\ & - \frac{1}{2\sigma_2^2(1-\delta^2)}\|(\boldsymbol{R} - \boldsymbol{M}B - \boldsymbol{X}\boldsymbol{\theta}_2) - \kappa(\boldsymbol{M} - \boldsymbol{X}\boldsymbol{\theta}_1)\|_2^2,\end{aligned} \tag{10}$$

where $\|\mathbf{x}\|_2$ is the $\ell_2$-norm of vector $\mathbf{x}$; $\boldsymbol{R} = (R_{p+1}, \ldots, R_T)^\top$, $\boldsymbol{M} = (M_{p+1}, \ldots, M_T)^\top$, $\boldsymbol{X} = (\boldsymbol{X}_{p+1}, \ldots, \boldsymbol{X}_T)^\top$; and $\kappa = \delta\sigma_2/\sigma_1$.

In our model, $\delta$ accounts for the effect of unmeasured confounding. For independent observations, it has been shown that $\delta$ is not identifiable with the existence of unmeasured confounding (Imai et al., 2010; Zhao and Luo, 2014). For time series data, we prove that this nonidentifiable issue cannot be alleviated in the single-level GMA either, and we will later show that it can be identified in our multilevel GMA model in Section 2.4.

THEOREM 1: *For every fixed $\delta \in (-1, 1)$, given the initial $p$ observations, $\ell(\boldsymbol{\theta}_1, \boldsymbol{\theta}_2, B, \boldsymbol{\Sigma})$ achieves the same maximum conditional likelihood value, where the maximum is taken over parameters $\{\boldsymbol{\theta}_1, \boldsymbol{\theta}_2, B, \sigma_1, \sigma_2\}$.*

Given $\delta$, the conditional maximum likelihood estimator (CMLE) of the remaining parameters, however, is given in explicit forms by the following theorem.

THEOREM 2: *Given the initial $p$ observations $(Z_1, M_1, R_1), \ldots, (Z_p, M_p, R_p)$, for a fixed $\delta$ value, the conditional maximum likelihood estimator of the rest parameters in models (5) and (6) are*

$$\begin{aligned}\hat{\boldsymbol{\theta}}_1 &= (\boldsymbol{X}^\top\boldsymbol{X})^{-1}\boldsymbol{X}^\top\boldsymbol{M}, \\ \hat{\boldsymbol{\theta}}_2 &= \left(\boldsymbol{X}^\top(\mathbf{I} - \boldsymbol{P}_M)\boldsymbol{X}\right)^{-1}\boldsymbol{X}^\top(\mathbf{I} - \boldsymbol{P}_M)\boldsymbol{R} + \hat{\kappa}(\boldsymbol{X}^\top\boldsymbol{X})^{-1}\boldsymbol{X}^\top\boldsymbol{M}, \\ \hat{B} &= (\boldsymbol{M}^\top\boldsymbol{M})^{-1}\boldsymbol{M}\left(\mathbf{I} - \boldsymbol{X}(\boldsymbol{X}^\top(\mathbf{I} - \boldsymbol{P}_M)\boldsymbol{X})^{-1}\boldsymbol{X}^\top(\mathbf{I} - \boldsymbol{P}_M)\right)\boldsymbol{R} - \hat{\kappa}, \\ \hat{\sigma}_1^2 &= \frac{1}{T-p}\boldsymbol{M}^\top(\mathbf{I} - \boldsymbol{P}_X)\boldsymbol{M}, \\ \hat{\sigma}_2^2 &= \frac{1}{(T-p)(1-\delta^2)}\boldsymbol{R}^\top(\mathbf{I} - \boldsymbol{P}_{MX} - \boldsymbol{P}_M)\boldsymbol{R},\end{aligned}$$

*where* $\hat{\kappa} = \delta\hat{\sigma}_2/\hat{\sigma}_1$*;* $\boldsymbol{P}_X = \boldsymbol{X}(\boldsymbol{X}^\top\boldsymbol{X})^{-1}\boldsymbol{X}^\top$*,* $\boldsymbol{P}_M = \boldsymbol{M}(\boldsymbol{M}^\top\boldsymbol{M})^{-1}\boldsymbol{M}^\top$*, and* $\boldsymbol{P}_{MX} = (\mathbf{I} - \boldsymbol{P}_M)\boldsymbol{X}(\boldsymbol{X}^\top(\mathbf{I} - \boldsymbol{P}_M)\boldsymbol{X})^{-1}\boldsymbol{X}^\top(\mathbf{I} - \boldsymbol{P}_M)$ *are projection matrices.*

From Theorem 2, the estimator of $\boldsymbol{\theta}_1$ is independent of $\delta$ since $Z_t$ is randomized. The estimators of $B$ and $\boldsymbol{\theta}_2$ contain additional terms that are functions of $\hat{\kappa}$ or $\delta$. These terms serve as bias correction due to unmeasured confounding.

Under the stationary condition, the impact of the initial $p$ observations is negligible when evaluating the likelihood (Akaike, 1973). We show that our estimator in Theorem 2 is consistent, and the asymptotic covariance matrix is derived in Theorem A.1.

We illustrate these theorems using a toy simulated data set. We assume a MAR(1) model for $(E_{1t}, E_{2t})$ with true $\delta = 0.5$. In Figure 4a, the (profile) conditional likelihood yields the same maximum value for varying $\delta$ as predicted by Theorem 1. Therefore, $\delta$ cannot be estimated by further maximizing this maximum (profile) conditional likelihood over $\delta$ alone. Alternatively, one may consider $\delta$ as a sensitivity parameter, and perform sensitivity analysis as $\delta$ varies. Figure 4b presents the estimates for $AB$ as $\delta$ varies, where the confidence intervals are calculated from the asymptotic formula (Theorem A.5). As shown in the figure, the estimated $AB$ value is sensitive to the choice of $\delta$. If assuming no unmeasured confounding ($\delta = 0$), the estimate of $AB$ even falls outside the confidence interval under the true $\delta$. This motivates us to consider an extension to model our two-level data jointly in the next section, although the single-level method without the extension can be applied to time series if unmeasured confounding is not a concern.

[Figure 4 about here.]

## 2.4 *Extension to two-level data*

In this section, we extend our GMA model to our two-level data, adapting the multilevel mediation method for independent data proposed by Zhao and Luo (2014).

2.4.1 *Model.* We will refer to the two levels as participant and scan time in this paper. For the time series of participant $i$ ($i = 1, \ldots, N$), we model the first-level scan-time data by our single level GMA models (1), (2) and (3) as

$$\begin{cases} M_{i_t} = Z_{i_t} A_i + E_{i_{1t}} \\ R_{i_t} = Z_{i_t} C_i + M_{i_t} B_i + E_{i_{2t}} \end{cases}, \tag{11}$$

$$\begin{cases} E_{i_{1t}} = \sum_{j=1}^{p} \omega_{i_{11_j}} E_{i_{1,t-j}} + \sum_{j=1}^{p} \omega_{i_{21_j}} E_{i_{2,t-j}} + \epsilon_{i_{1t}} \\ E_{i_{2t}} = \sum_{j=1}^{p} \omega_{i_{12_j}} E_{i_{1,t-j}} + \sum_{j=1}^{p} \omega_{i_{22_j}} E_{i_{2,t-j}} + \epsilon_{i_{2t}} \end{cases},$$

where

$$\begin{pmatrix} \epsilon_{i_{1t}} \\ \epsilon_{i_{2t}} \end{pmatrix} \sim \mathcal{N}\left(\mathbf{0}, \mathbf{\Sigma}_i\right), \quad \mathbf{\Sigma}_i = \begin{pmatrix} \sigma_{i_1}^2 & \delta_i \sigma_{i_1} \sigma_{i_2} \\ \delta_i \sigma_{i_1} \sigma_{i_2} & \sigma_{i_2}^2 \end{pmatrix}; \tag{12}$$

$A_i$, $B_i$ and $C_i$ are the model coefficients of participant $i$. In order to estimate the population level causal effects after accounting for the between-participant variation, we employ the following multivariate linear model

$$b_i = \begin{pmatrix} A_i \\ B_i \\ C_i \end{pmatrix} = \begin{pmatrix} A \\ B \\ C \end{pmatrix} + \begin{pmatrix} \epsilon_i^A \\ \epsilon_i^B \\ \epsilon_i^C \end{pmatrix} = b + \eta_i, \tag{13}$$

where $A$, $B$ and $C$ denote the population level coefficients; and $\epsilon_i^A$, $\epsilon_i^B$ and $\epsilon_i^C$ are the random errors of participant $i$, which are assumed to be identically distributed from a trivariate normal distribution with mean zero and covariance matrix $\mathbf{\Lambda}$. The population direct effect is $C$, and the population indirect effect is $AB$ by the product method. There is an alternative definition of the population indirect effect by the difference method, see the discussion in Kenny et al. (2003). This approach would also require fitting a total effect model by regressing outcome $R$ on treatment $Z$ only for each participant, and a population equation analogous to model (13). For the sake of space, we omit the description of this alternative approach in this paper, because they yield very similar numerical results for the indirect effect.

As discussed in Section 2.3, we estimate the parameters in (11) through the transformed model,

$$\begin{aligned} M_{i_t} &= Z_{i_t} A_i + \sum_{j=1}^{p} \phi_{i_{1j}} Z_{i_{t-j}} + \sum_{j=1}^{p} \psi_{i_{11_j}} M_{i_{t-j}} + \sum_{j=1}^{p} \psi_{i_{21_j}} R_{i_{t-j}} + \epsilon_{i_{1t}} \\ &= \boldsymbol{X}_{i_t}^{\top} \boldsymbol{\theta}_{i_1} + \epsilon_{i_{1t}}, \end{aligned} \tag{14}$$

$$\begin{aligned} R_{i_t} &= Z_{i_t} C_i + M_{i_t} B_i + \sum_{j=1}^{p} \phi_{i_{2j}} Z_{i_{t-j}} + \sum_{j=1}^{p} \psi_{i_{12_j}} M_{i_{t-j}} + \sum_{j=1}^{p} \psi_{i_{22_j}} R_{i_{t-j}} + \epsilon_{i_{2t}} \\ &= M_{i_t} B_i + \boldsymbol{X}_{i_t}^{\top} \boldsymbol{\theta}_{i_2} + \epsilon_{i_{2t}}, \end{aligned} \tag{15}$$

where $\boldsymbol{X}_{i_t}$, $\boldsymbol{\theta}_{i_1}$ and $\boldsymbol{\theta}_{i_2}$ are defined the same as in Section 2.3 for participant $i$, $i = 1, \ldots, N$.

As shown in Theorem 1, $\delta_i$ is not identifiable from the likelihood function for each participant $i$, and thus it is easy to see that one cannot identify different $\delta_i$ from the overall likelihood function of $N$ independent participants. To identify $\delta_i$ from data, we adapt the optimization methods in Zhao and Luo (2014), and we need to impose the following assumption.

(A6) $\delta_i$ is constant across participants, i.e., $\delta_i = \delta$ for all $i$.

It is also worth noting that one alternative proposal is to perform sensitivity analysis using different $\delta_i$ for each $i$ without assuming (A6). However, the number of sensitivity parameters makes this proposal unrealistic for large $N$. Under (A6), we will introduce in the next section two data-driven methods to estimate $\delta$ by pooling data across participants.

2.4.2 *Method.* The principal idea in Zhao and Luo (2014) is to estimate $\delta$ by maximizing the joint likelihood of $N$ participants. We adapt this idea for our GMA model here. Let $\boldsymbol{\Upsilon} = (\delta, b, \boldsymbol{\Lambda}, (\boldsymbol{\theta}_{i_1}, \boldsymbol{\theta}_{i_2}, B_i), (\sigma_{1_i}, \sigma_{2_i}))$, the likelihood function (conditional on the initial $p$ observations) is written as

$$
\begin{aligned}
h(\boldsymbol{\Upsilon}) =& \sum_{i=1}^{N}\sum_{t=p+1}^{T_i} \log \mathbb{P}\left(R_{i_t}, M_{i_t} \mid Z_{i_t}, \boldsymbol{Z}_{i_{t-1}}^{(p)}, \boldsymbol{M}_{i_{t-1}}^{(p)}, \boldsymbol{R}_{i_{t-1}}^{(p)}, \boldsymbol{\theta}_{i_1}, \boldsymbol{\theta}_{i_2}, B_i, \delta, \sigma_{i_1}, \sigma_{i_2}\right) \\
&+ \sum_{i=1}^{N} \log \mathbb{P}\left(b_i \mid b, \boldsymbol{\Lambda}\right) \\
=& h_1 + h_2,
\end{aligned}
\tag{16}
$$

where $b_i = (A_i, B_i, C_i)$, $A_i$ and $C_i$ are the first element of $\boldsymbol{\theta}_{i_1}$ and $\boldsymbol{\theta}_{i_2}$, respectively; $h_1$ is the sum of $N$ log-likelihood functions (10), and $h_2$ is the log-likelihood function of model (13). It is challenging to optimize these many parameters that grow with $N$. In particular, our GMA model contains more parameters than the previous multilevel mediation model, since we include those temporal dependence parameters as well. We thus propose two algorithms for maximizing the joint likelihood, with different computational complexity and numerical accuracy.

**A two-stage algorithm**. This algorithm is inspired by the two-level GLM method common for fMRI analysis, see for example Kenny et al. (2003) and Lindquist (2008). In the first stage, we estimate, for each participant $i$, the coefficients in the single level model with a given $\delta$ using Theorem 2. This stage splits the computation cost by maximizing the summands in $h_1$ for each participant, and can be computed in parallel. In the second stage, we plug in the estimated coefficients from the first stage into the left-hand side of the second level regression model (13), and we can easily maximize its likelihood function $h_2$. To identify $\delta$, we repeat the two-stage computation for different $\delta$ while using a one-dimensional optimization algorithm (e.g. Newton's method) to find the $\delta$ that yields the maximum joint likelihood $h$.

The key challenge for proving the consistency of this algorithm is to show that $\delta$ is identifiable and is estimated consistently using the above algorithm, as the consistency of the remaining parameters (given $\delta$) are guaranteed by the standard maximum likelihood theory under regularity conditions. For example, the consistency of the first level estimates is given in Theorem A.1 of the supplementary materials. We address this challenge in the following theorem.

THEOREM 3: *Assume assumptions (A1)-(A6) are satisfied, and the stationary condition of the single level model holds. Assume* $\mathbb{E}(Z_{i_t}^2) = q < \infty$*, for* $i = 1, \ldots, N$*. Let* $T = \min_i T_i$*.*

*(a) If* $\mathbf{\Lambda}$ *is known, then the two-stage estimator* $\hat{\hat{\delta}}$ *maximizes the profile likelihood of model* (13) *asymptotically, and* $\hat{\hat{\delta}}$ *is* $\sqrt{NT}$*-consistent.*

*(b) If* $\mathbf{\Lambda}$ *is unknown, then the profile likelihood of model* (13) *has a unique maximizer* $\hat{\hat{\delta}}$ *asymptotically, and* $\hat{\hat{\delta}}$ *is* $\sqrt{NT}$*-consistent, provided that* $1/\varpi = \bar{\kappa}^2/\varrho^2 = \mathcal{O}_p(1/\sqrt{NT})$*,* $\kappa_i = \sigma_{i_2}/\sigma_{i_1}$*,* $\bar{\kappa} = (1/N)\sum \kappa_i$*, and* $\varrho^2 = (1/N)\sum(\kappa_i - \bar{\kappa})^2$*.*

Based on the consistency for $\delta$, we prove the consistency of all other parameters in the following corollary.

COROLLARY 1: *Under assumptions in Theorems 2, A.1 and 3, the estimators of model coefficients introduced in Theorem 2 are consistent with asymptotic joint distribution as in Theorem A.1. Further, the estimator of population-level parameters in model* (13) *are consistent.*

To verify the identifiability in practice, we plot the maximum log-likelihood value against $\delta$. Figure 5a illustrates such a plot for the toy simulated data set analyzed before. The likelihood is unimodal, while the single level likelihood in Figure 4a is flat. The two-stage estimator of $\delta$ at the peak is 0.466 with the true $\delta = 0.5$.

**A block coordinate-descent algorithm**. Though the two-stage algorithm is computationally light and asymptotically consistent, it only approximately maximizes the joint likelihood $h$. To improve the finite sample performance, we propose a block coordinate-descent algorithm for better maximizing $h$. Some finite sample improvement was observed by a similar strategy in Zhao and Luo (2014).

**Algorithm 2.1** A block coordinate-descent algorithm to compute all other parameters given $\delta$ and estimate $\delta$ in our two-level model using likelihood function (16).

**Compute the likelihood value and coefficient estimates for a given $\delta$:**

(1) Estimate $(\sigma_{i_1}, \sigma_{i_2})$, $(\boldsymbol{\theta}_{i_1}, \boldsymbol{\theta}_{i_2}, B_i)$, $b$ and $\boldsymbol{\Lambda}$ by maximizing the likelihood (16) over these remaining parameters using block coordinate descent.

(2) Return the maximum likelihood value.

**When $\delta$ is unknown, apply an optimization algorithm (e.g., Newton's method) to maximize over $\delta$ using the maximum likelihood value at Step 2.**

We formally formulate the following optimization problem

$$\max_{\boldsymbol{\Upsilon}:\left((\sigma_{i_1}, \sigma_{i_2}), \boldsymbol{\Lambda}\right) \in \mathcal{S}} h(\boldsymbol{\Upsilon}), \tag{17}$$

where $\mathcal{S}$ is a constraint set for the variance components. We put a positive constraint on each $(\sigma_{i_1}, \sigma_{i_2})$, and a positive definite constraint on $\boldsymbol{\Lambda}$. We divide these many variables into blocks based on the following theorem.

THEOREM 4: *Assume $\delta$ is given. The negative of likelihood function (16) is conditional convex in the parameter sets $(\sigma_{i_1}^{-1}, \sigma_{i_2}^{-1})$, $(\boldsymbol{\theta}_{i_1}, \boldsymbol{\theta}_{i_2}, B_i)$, $b$ and $\boldsymbol{\Lambda}^{-1}$, respectively. The conditional optimizer for each parameter set is given in explicit forms in Section A.9.*

For a given $\delta$, each block/set of variables in the theorem are given in explicit forms for each iterative update, which eases the computation. As before, we estimate $\delta$ by a one-dimensional optimization algorithm. The full algorithm is summarized in Algorithm 2.1.

We propose to check the solution of $\delta$ graphically as before. Figure 5b presents the likelihood $h$ for our toy simulated data set. It is a unimodal function of $\delta$ peaked at our block-coordinate algorithm estimate $\hat{\delta} = 0.492$, which is closer to the truth 0.5 in this numerical example.

[Figure 5 about here.]

### 2.5 *Inference*

Because the distribution of the product $\hat{A}\hat{B}$ can be far from Gaussian, we propose to employ bootstrap over participants to perform statistical inference on the population causal effects.

## 3. Simulation Study

We compare our GMA estimators with other methods on simulated data in this section. We simulate both single-level data and two-level data, with and without unmeasured confounding, to assess the performance under different scenarios. For single-level mediation models, we consider the GMA (GMA-$\delta$) estimator in Section 2.3 with the true $\delta$ given, the GMA estimator (GMA-0) assuming no unmeasured confounding or $\delta = 0$, the method for independent observations proposed by Zhao and Luo (2014) (MACC-$\delta$, implemented using R macc package) with also the true $\delta$, and the Baron-Kenny (BK) method (Baron and Kenny, 1986). The MACC-$\delta$ method is shown to correct the bias due to $\delta$, and has similar performance with other cause mediation methods (e.g. Imai et al. (2010)) when $\delta = 0$. For two-level mediation models, we consider our two-stage (GMA-ts) and block coordinate-descent (GMA-h) approaches in Section 2.4.2, the two-level methods in Zhao and Luo (2014) (MACC-h and MACC-ts), the multilevel SEM method proposed by Kenny et al. (2003) (KKB), and the BK method. All other non-GMA methods considered here are developed for independent observations. GMA-h, GMA-ts, MACC-h and MACC-ts are developed to estimate $\delta$ from data, and thus correct for unmeasured confounding, while KKB and BK are not.

### 3.1 *Single level mediation analysis*

We simulate data sets of 100 time points from the following models. At each time point $t$, $Z_t$ is generated from a Bernoulli distribution with probability 0.5 to be one. The null hypothesis for the direct effect is $\mathrm{H}_0 : C = 0$; the null hypothesis for the indirect effect under the product definition is $\mathrm{H}_0 : AB = 0$, where at least one of $A$ and $B$ is zero. We here present the results when $B = 0$ and $A \neq 0$. Under the alternative, we set the coefficients as $A = 0.5$, $B = 1$, and $C = 0.5$. For simplicity, in this simulation study, we consider a MAR(1) model for $(E_{1t}, E_{2t})$. The marginal variances of the Gaussian white noise $(\epsilon_{1t}, \epsilon_{2t})$ are $\sigma_1 = 1$ and $\sigma_2 = 2$, respectively. The correlation $\delta$ is set to be either 0.5 or 0. The initial joint distribution of $(E_{10}, E_{20})$ is assumed to follow a bivariate normal distribution with mean zero and covariance $2\boldsymbol{\Sigma}$. To make the stationary condition satisfied, i.e.,

$$2\boldsymbol{\Sigma} = \boldsymbol{\Omega}^\top (2\boldsymbol{\Sigma})\,\boldsymbol{\Omega} + \boldsymbol{\Sigma}, \tag{18}$$

where $\boldsymbol{\Omega}$ is the transition matrix, the elements in $\boldsymbol{\Omega}$ are specified as

$$\boldsymbol{\Omega} = \begin{pmatrix} \omega_{11} & \omega_{12} \\ \omega_{21} & \omega_{22} \end{pmatrix} = \begin{pmatrix} -0.809 & -0.618 \\ 0.154 & -0.500 \end{pmatrix} \text{ (when } \delta = 0.5);$$

$$\boldsymbol{\Omega} = \begin{pmatrix} -0.5 & -1 \\ 0.25 & -0.5 \end{pmatrix} \text{ (when } \delta = 0).$$

For each simulation setting, we drop the first 1000 samples from the model as the burn-in samples, in order to stabilize the time series. All simulations are repeated 1000 times.

We compare the estimates of the direct effect $C$, the indirect effect $AB$, the variances $\sigma_1^2$ and $\sigma_2^2$, and the transition matrix $\boldsymbol{\Omega}$ in Table 1.

[Table 1 about here.]

For all the settings, both GMA-$\delta$ and MACC-$\delta$ yield estimates of $C$ and $AB$ close to the true values, but GMA-$\delta$ yields closer variance estimates for $\sigma_1^2$ and $\sigma_2^2$. For the GMA-$\delta$ method, Theorem A.1 and Theorem A.5 guarantee its asymptotic unbiasedness. MACC-$\delta$ also accounts for the unmeasured confounding effect but treats each time point as an independent sample. Though ignoring the temporal correlations by MACC-$\delta$ does not introduce asymptotic bias similar to the unbiasedness of the ordinary least squares estimator for time series (see for example Theorem 3.11 of Shao (2003)). The variance estimates for $\sigma_1^2$ and $\sigma_2^2$ are biased and thus the asymptotic variance formulas cannot be used for inference. The difference in the asymptotic variance estimates are due to the MAR parameters, see our Theorem A.1. Due to the difference in variance estimates, GMA-$\delta$ has more statistical power as seen the simulations results. Both GMA-0 and BK fail to account for unmeasured confounding and time series correlations, they both yield biased estimates for the causal effects and variances.

Our GMA methods also yield estimates for the transition matrix $\mathbf{\Omega}$. GMA-$\delta$ yields estimates close to the truth across all scenarios while GMA-0 introduces biases for some elements in $\mathbf{\Omega}$ when $\delta \neq 0$. The reason is that $\delta$ influences the transition parameter estimates, as predicted by our Theorem 2.

3.2 *Two-level mediation analysis*

We simulate two-level mediation data from the following settings. The total number of participant is $N = 50$. For each participant, the number of time points (trials) is a random draw from the Poisson distribution with mean 100, and the length of burn-ins is 2000. The population-level coefficients are set to be $A = 0.5$, $B = -1$ and $C = 0.5$. The errors in the coefficient regression model (13) are assumed to be normally distributed with mean zero and covariance matrix $\mathbf{\Lambda} = \text{diag}\{0.5, 0.5, 0.5\}$. For each participant, similar to the single level simulation setting, we assume a MAR(1) model for $(E_{i_1t}, E_{i_2t})$. The marginal variance of the Gaussian white noise $(\epsilon_{i_1t}, \epsilon_{i_2t})$ are $\sigma_{i_1} = \sigma_1 = 1$ and $\sigma_{i_2} = \sigma_2 = 2$ (for $i = 1, \ldots, N$), respectively. We consider varying $\delta$ in our simulations. The initial joint distribution of $(E_{i_10}, E_{i_20})$ is assumed to be a bivariate normal distribution with mean zero and covariance $2\mathbf{\Sigma}$ (for $i = 1, \ldots, N$), and the transition matrix $\mathbf{\Omega}$ is calculated to satisfy the stationary condition for each $\delta$ value. The simulations are repeated 200 times.

We first compare the mediation effect estimates under different $\delta$ values (Figure 6). From the figure, GMA-h has the smallest bias overall. GMA-ts and the two MACC methods also yield small bias, while the KKB and BK estimates have large and increasing bias as $|\delta|$ increases. This shows that GMA-h improves GMA-ts in finite samples. Though the MACC methods are developed to correct for nonzero $\delta$, they yield larger bias here partly because the likelihood formulation (Zhao and Luo, 2014) is misspecified for time series data without accounting for the temporal correlations. KKB and BK cannot correct for the unmeasured confounding factor $\delta$, and thus their estimates are sensitive to $\delta$.

In Figure 6e, we present the GMA-h estimate of the transition matrix $\mathbf{\Omega}$. The estimates are averaged over all participants to simplify the plot. The transition matrix estimates are close to the truth, which demonstrates that our GMA-h method produces low bias after correcting for the confounding effect due to nonzero $\delta$.

[Figure 6 about here.]

We also consider increasing sample sizes to validate the consistency theory in Theorem 3. For a fixed $\delta = 0.5$, we consider sample size $N = T_i = 50, 200, 500, 1000, 5000$. Figure 7 shows that the GMA-ts estimates of $\delta$ and $AB$ converge to the true value as $N$ and $T_i$ increase. GMA-h shows a slight improvement over GMA-ts when $N < 200$, but the differences become negligible for larger $N$. Finally, we also present the estimate of $\mathbf{\Omega}$ from GMA-h method in Figures 7c and 7f, which shows that the estimates of all elements in $\mathbf{\Omega}$ converge to the true values as $N$ increases.

[Figure 7 about here.]

## 4. The fMRI Experiment

The data set was obtained from the OpenfMRI database, and the accession number is ds000030. In the experiment, $N = 121$ right-handed participants in healthy condition were recruited. The participants were asked to perform motor responses to two types of randomized stimuli: GO or STOP. The STOP/GO stimuli were randomly intermixed with 96 GO and 32 STOP stimuli, with randomly jittered time intervals between the stimuli. Under the GO stimulus, the participants should respond with button presses, when seeing the stimuli; under the STOP stimulus, the participant should withhold from pressing when a stop signal (a 500 Hz tone presented through headphones) was presented after the GO stimulus. Each participant was scanned using Siemens Trio 3T scanners. The experiment was about six minutes with 128 trials for each participant. For each participant, 184 images were acquired (34 slices with slice thickness 4 mm, TR = 2 s, TE = 30 ms, flip angle = 90°, matrix $64 \times 64$, and FOV = 192 mm, oblique slice orientation). More details about the experiment can be found in Poldrack et al. (2016).

The preprocessing for both anatomical and functional images was conducted using Statistical Parametric Mapping version 5 (SPM5) (Wellcome Department of Imaging Neuroscience, University College London, London, UK), including slice timing correction, realignment, coregistration, normalization, and smoothing. The BOLD series of $T_i = T = 184$ time points were extracted by averaging over the voxels within 10 mm radius sphere centered at preSMA (MNI coordinate: $(-4, -8, 60)$) and M1 (MNI coordinate: $(-41, -20, 62)$). Our main objective is to study how the preSMA activity mediates the STOP/GO stimulus effect on the M1 activity, as well as how these two brain regions are temporally related. Following the standard SPM approach to model the BOLD response, we convolute the stimulus sequence with the Canonical hemodynamic response function (HRF) (Friston, 2009) to generate the treatment time series.

We compare the mediation effect estimates from the proposed GMA-h and GMA-ts methods with the MACC-h, KKB and BK methods. Because other competing methods do not provide estimates for the transition matrix, we also compare the transition matrix estimates with the MAR fits by Harrison et al. (2003), which does not model the mediation effects. It is usually sufficient to use AR(1) or AR(2) to model the temporal correlations in fMRI (Lindquist, 2008). Thus, we set the lag parameter $p = 2$ in our GMA approach. We also analyze our data using $p = 3$ but the lag-three temporal correlation estimates are close to zero (Section C.2 of the supplementary materials). All methods use 200 bootstrap samples for inference.

[Table 2 about here.]

Table 2 presents the estimates (and the 95% bootstrap confidence intervals) of $\delta$, direct effect $C$, and indirect effect $AB$. The estimate of $\delta$ from our GMA methods are negative and significant from zero, which provides the evidence of the existence of unmeasured confounding in the data. The estimates for $\delta$ by GMA-ts and GMA-h are close in this data set. Consistent with the simulation results, MACC-h produces a larger estimate for $\delta$.

Our GMA-ts and GMA-h also yield similar estimates for $C$ and $AB$, different from all other methods. In particular, our GMA methods yield the largest indirect effect estimates in magnitude. MACC-h yields a much smaller estimate (about 30% less) in magnitude. KKB and BK yield similar and the smallest estimates, because they fail to account for the confounding effect due to nonzero $\delta$. Though all these estimates give the same qualitative interpretation for the role of preSMA, which is consistent with the current scientific understanding, our quantitative estimates suggest a much larger role of preSMA than other methods.

[Figure 8 about here.]

Another advantage of our GMA methods is that it also estimates the temporal dependencies between two brain regions, which are represented by the transition matrix $\mathbf{\Omega}$. The bootstrap estimate for $\mathbf{\Omega}$ is shown in Figure 8a, where we observe significant feedback effects from M1 to preSMA at both lag levels ($\hat{\omega}_{21_1} = 0.100$ and $\hat{\omega}_{21_2} = -0.077$). Comparing with the estimates by the MAR method (see Figure 8b), we find that it, without modeling the direct and indirect effects like ours, produces larger point estimates for the diagonals and have larger bootstrap variability overall. It also yields wider confidence intervals for the off-diagonals than ours, though the point estimates are similar to ours.

## 5. Discussion

In this paper, we propose a mediation analysis framework for time series data. Our approach integrates multivariate autoregressive models and mediation analysis to yield a better understanding of such data. Our approach is also embedded in a causal mediation model for correlated errors, in order to address the unmeasured confounding issue. We prove that a simple two-stage algorithm will yield asymptotically unique and consistent estimates, and its finite sample performance is improved by a more sophisticated optimization algorithm with increased computational cost. Using both simulations and a real fMRI data set, we demonstrate the numerical advantages of our proposal.

Our model setup is motivated by several important statistical models for task-related fMRI data. It is likely that other scientific experiments or studies will require different modeling components, using other types of data with different data structures for the treatment, mediator and outcome. For example, some variables are scalars at the participant level, instead of time series, see the discussion of various multilevel data sets in Kenny et al. (2003). Time series modeling is also a topic with a long history, and some other time series models, other than multivariate autoregressive models, may be more suitable for certain experiments. We will leave exploring these different settings to future research.

We focus on randomized treatment in this paper, and this may not hold in observational studies. It is interesting to further develop our proposal using the tools for observational studies to relax the randomization requirement.

Many extensions of mediation models have also been considered in the literature, see VanderWeele (2015). These models can also include interactions and covariates, which are common in many social studies. We are interested in extending our proposal to these more complicated settings in the future.

**Acknowledgements**

This work was partially supported by National Institutes of Health grants R01EB022911, P20GM103645, P01AA019072, and S10OD016366, National Science Foundation grant DMS 1557467.

**Supplementary Materials**

Web Appendix is available with this paper at the Biometrics website on Wiley Online Library.

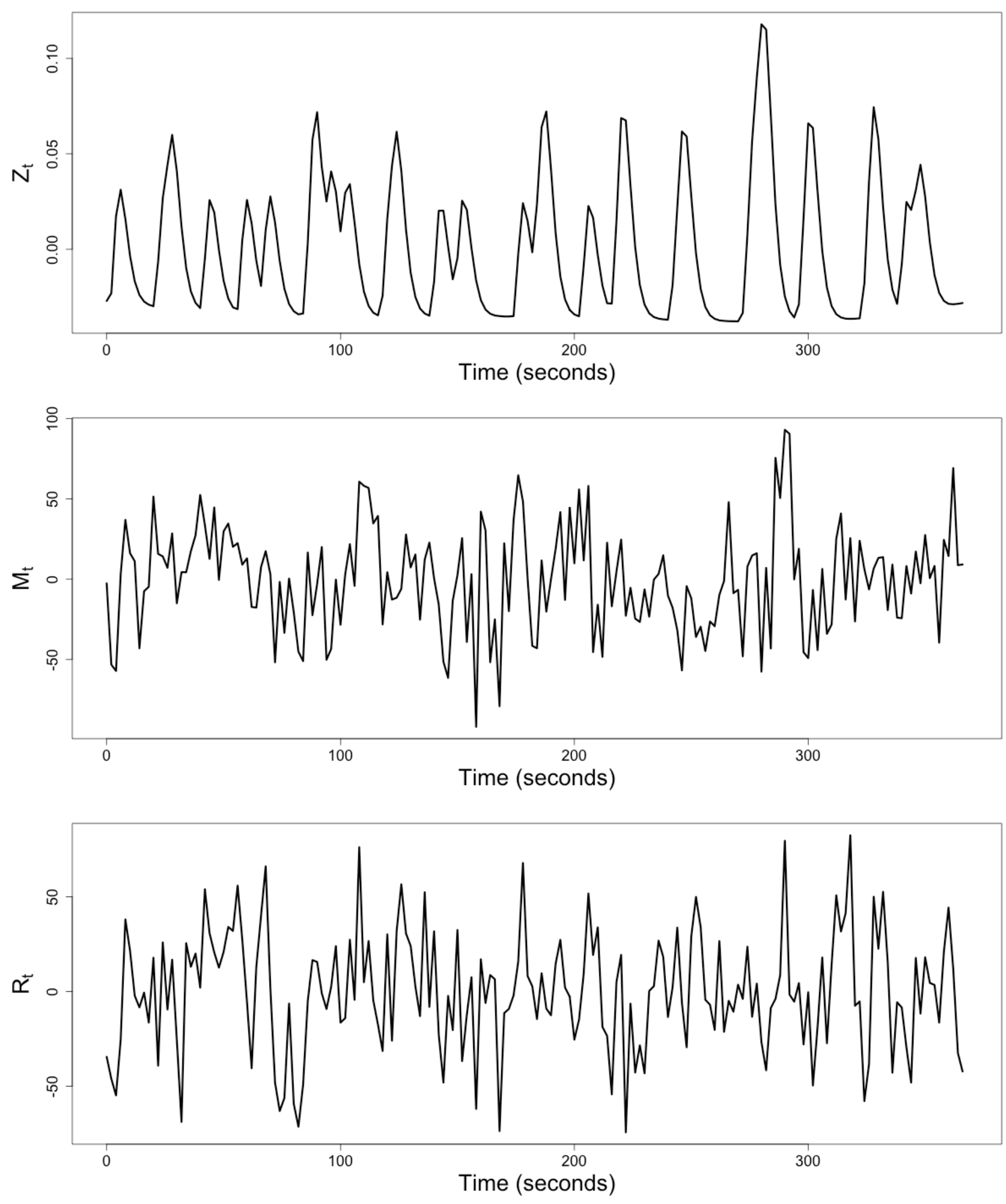

**Figure 1**: The stimulus input series ($Z_t$), and the preSMA ($M_t$) and M1 ($R_t$) fMRI BOLD time series from one of the 121 participants.

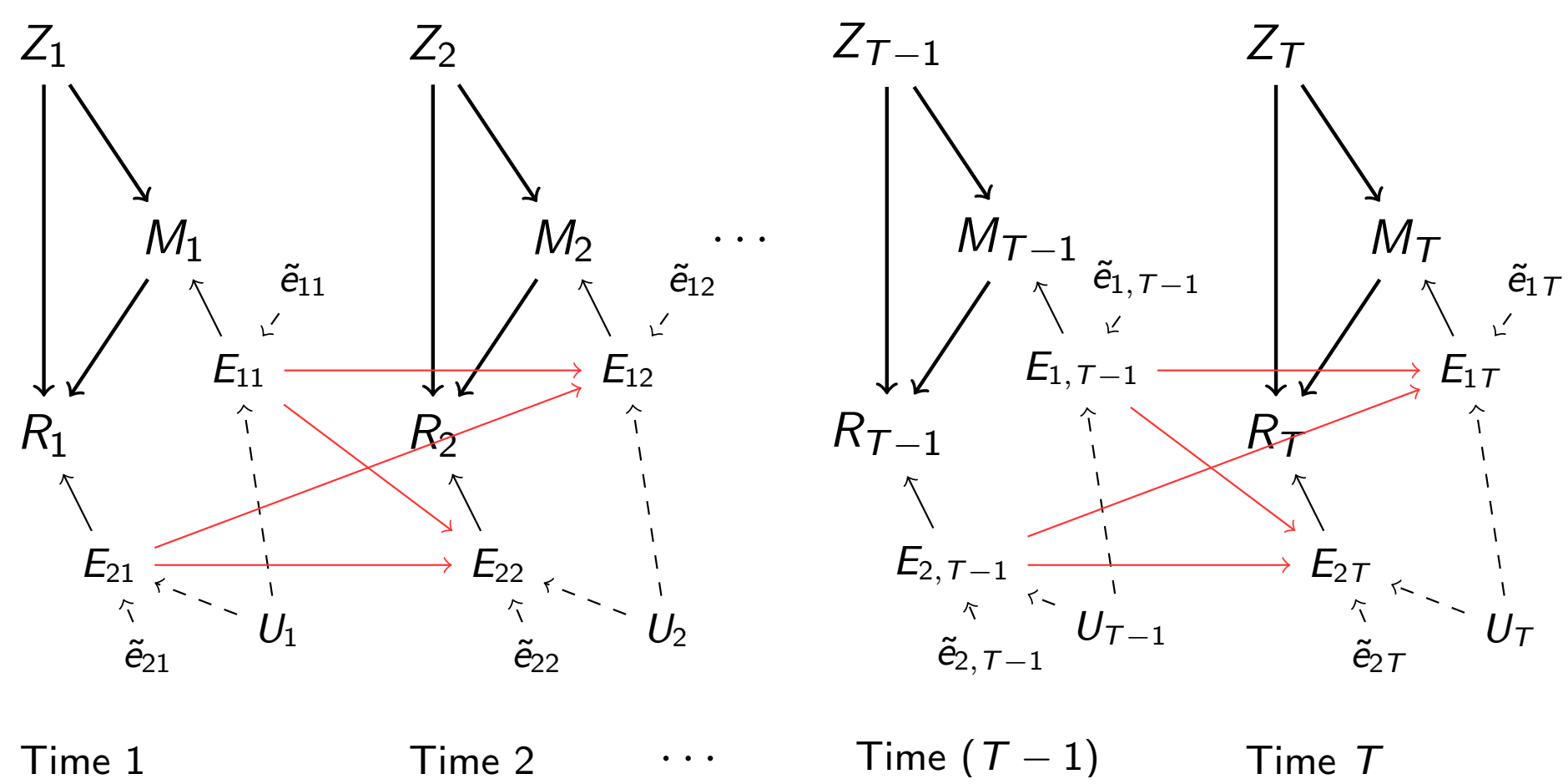

**Figure 2**: A conceptual diagram of the first level mediation model for time series data. At each time point $t$ $(t = 1, \ldots, T)$, the task-related signals are overlayed with the random fluctuations, $E_{1t}$ and $E_{2t}$, between which there exist both interregional and temporal dependencies. $U_t$ denotes the unmeasured confounding effect influencing both $E_{1t}$ and $E_{2t}$; $\tilde{e}_{1t}$ and $\tilde{e}_{2t}$ are independent errors, which are also independent of $U_t$, $t = 1, \ldots, T$.

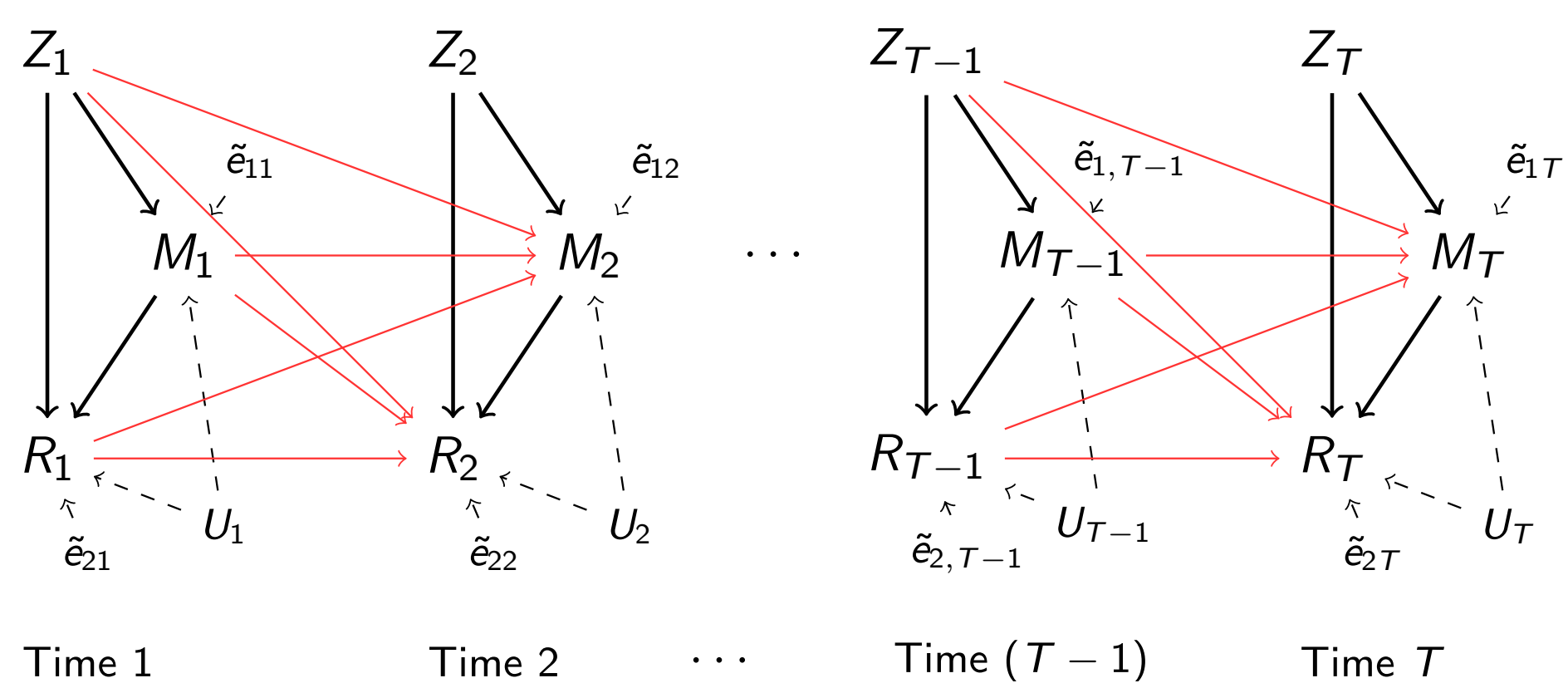

**Figure 3**: A conceptual diagram of models (5) and (6) when $p = 1$. At each time point $t$ ($t = 1, \ldots, T$), there is an unmeasured confounding variable $U_t$ influencing both $M_t$ and $R_t$. Meanwhile, both the treatment and the outcomes of the peovious time point $(Z_{t-1}, M_{t-1}, R_{t-1})$ have an impact on the current $M_t$ and $R_t$.

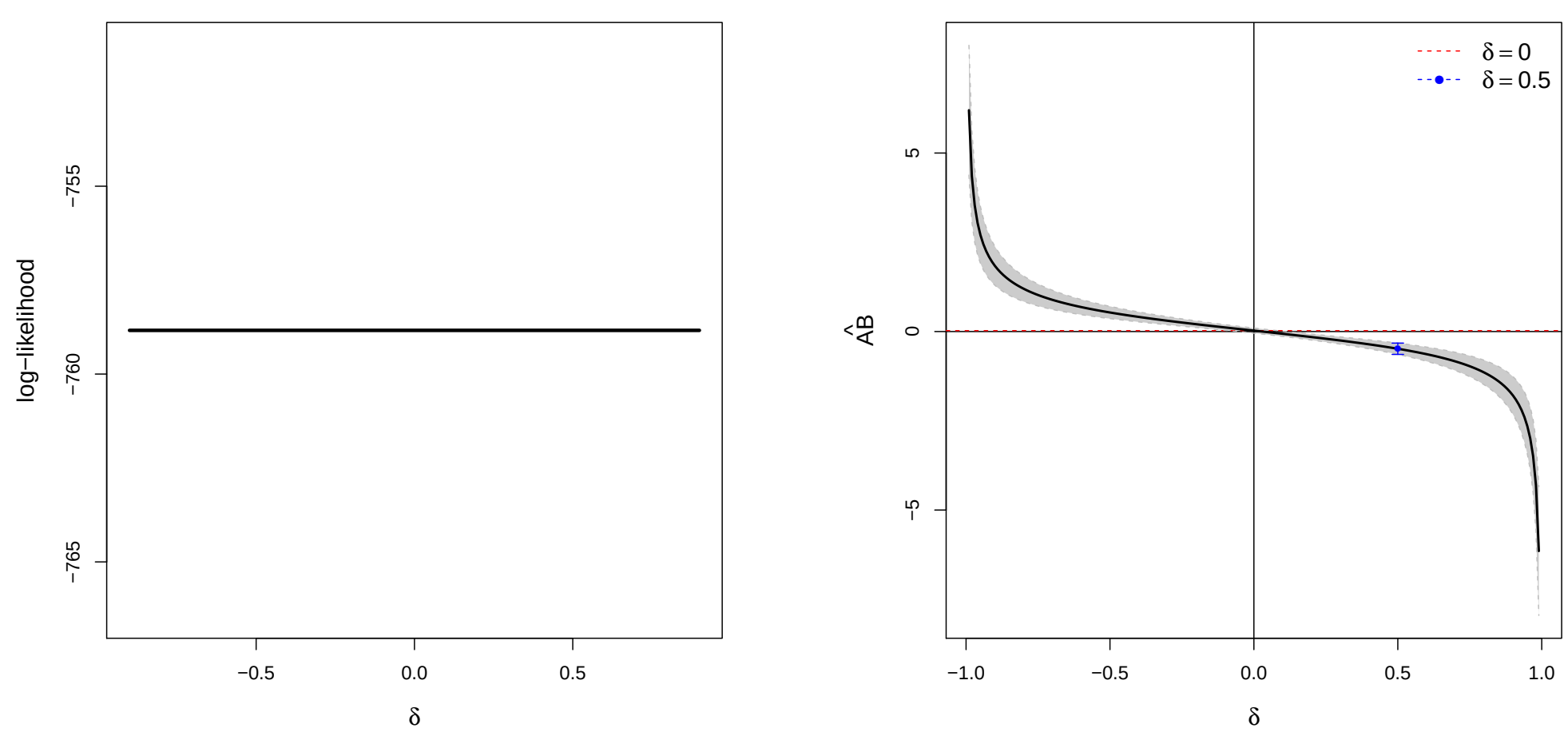

(a) Conditional likelihood of our single level model. (b) Sensitivity analysis of our single level model.

**Figure 4**: (a) Conditional log-likelihood and (b) sensitivity analysis of the indirect effect $(AB)$ under our single level GMA model with $p = 1$.

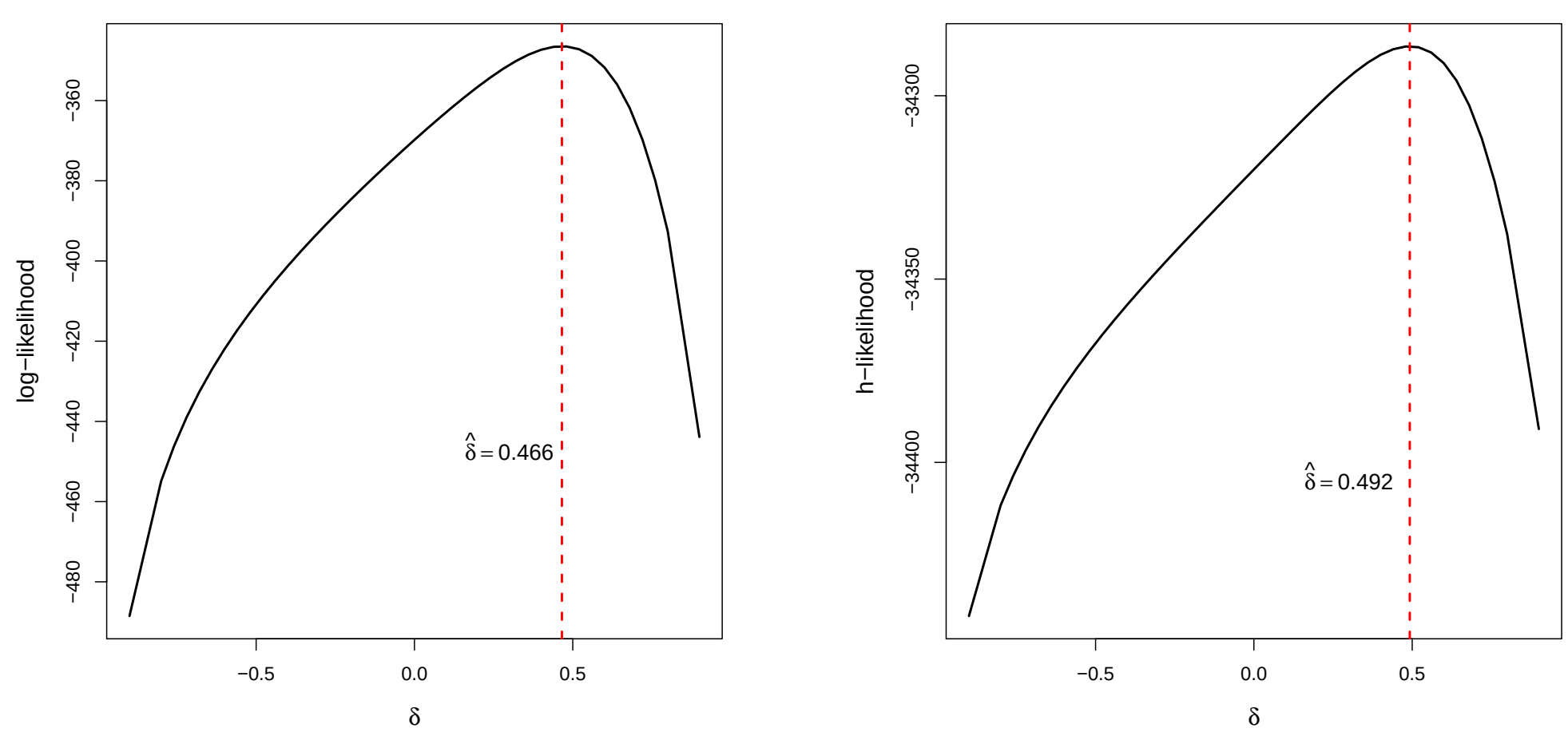

(a) Profile log-likelihood of the second stage regression model using the two-stage algorithm.

(b) Profile log-likelihood of the two-level model using the block coordinate-descent algorithm.

**Figure 5**: The log-likelihood functions under the two-level models of a simulated data set. The true $\delta$ value is 0.5. The red dashed line is the estimate from the two-stage and block coordinate-descent algorithms.

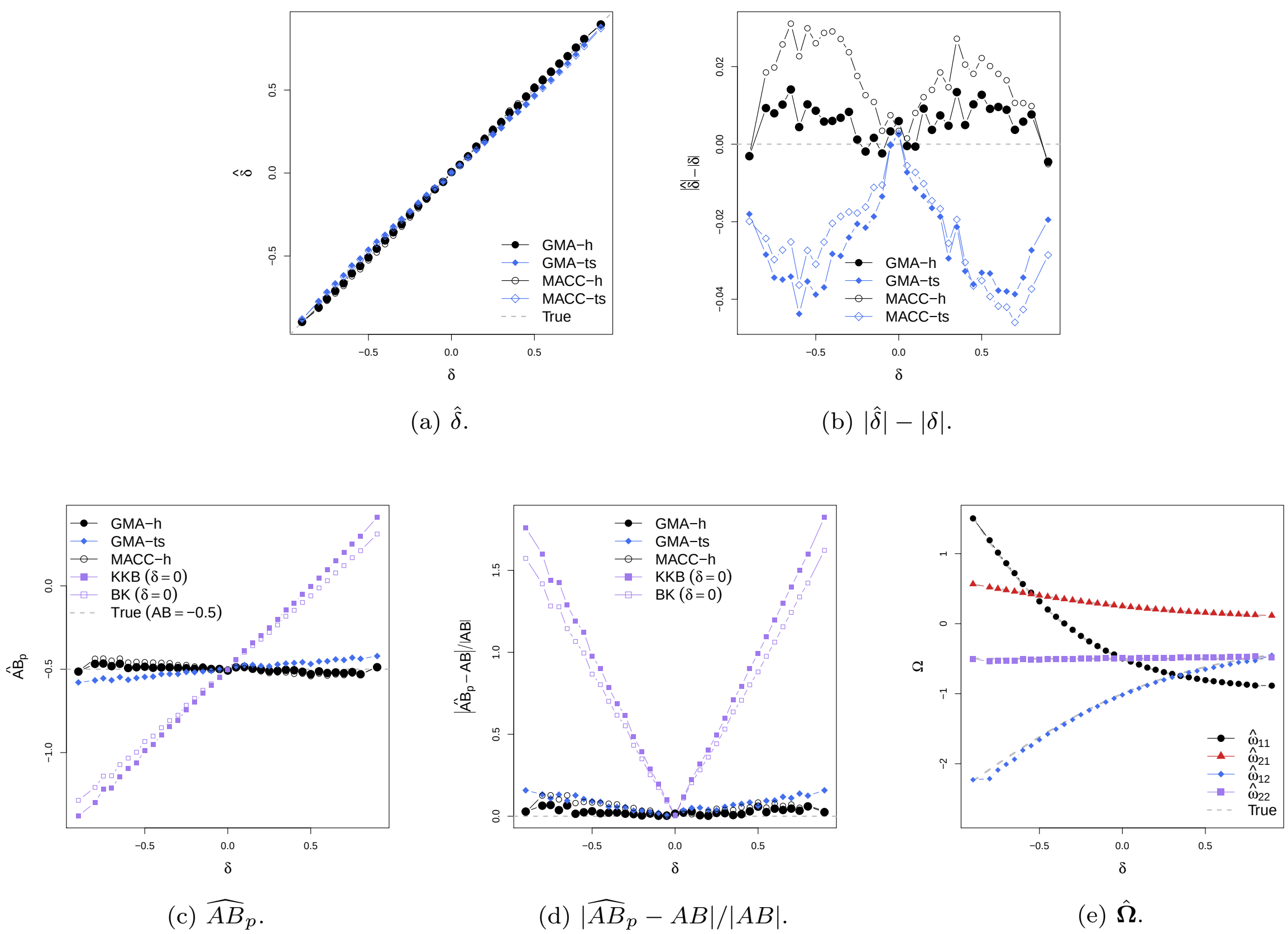

(a) $\hat{\delta}$. (b) $|\hat{\delta}| - |\delta|$.

(c) $\widehat{AB}_p$. (d) $|\widehat{AB}_p - AB|/|AB|$. (e) $\hat{\boldsymbol{\Omega}}$.

**Figure 6**: Point estimates and bias of $\delta$, the product estimates and the relative bias of $AB$, and point estimates of the transition matrix $\boldsymbol{\Omega}$ under different true $\delta$ values.

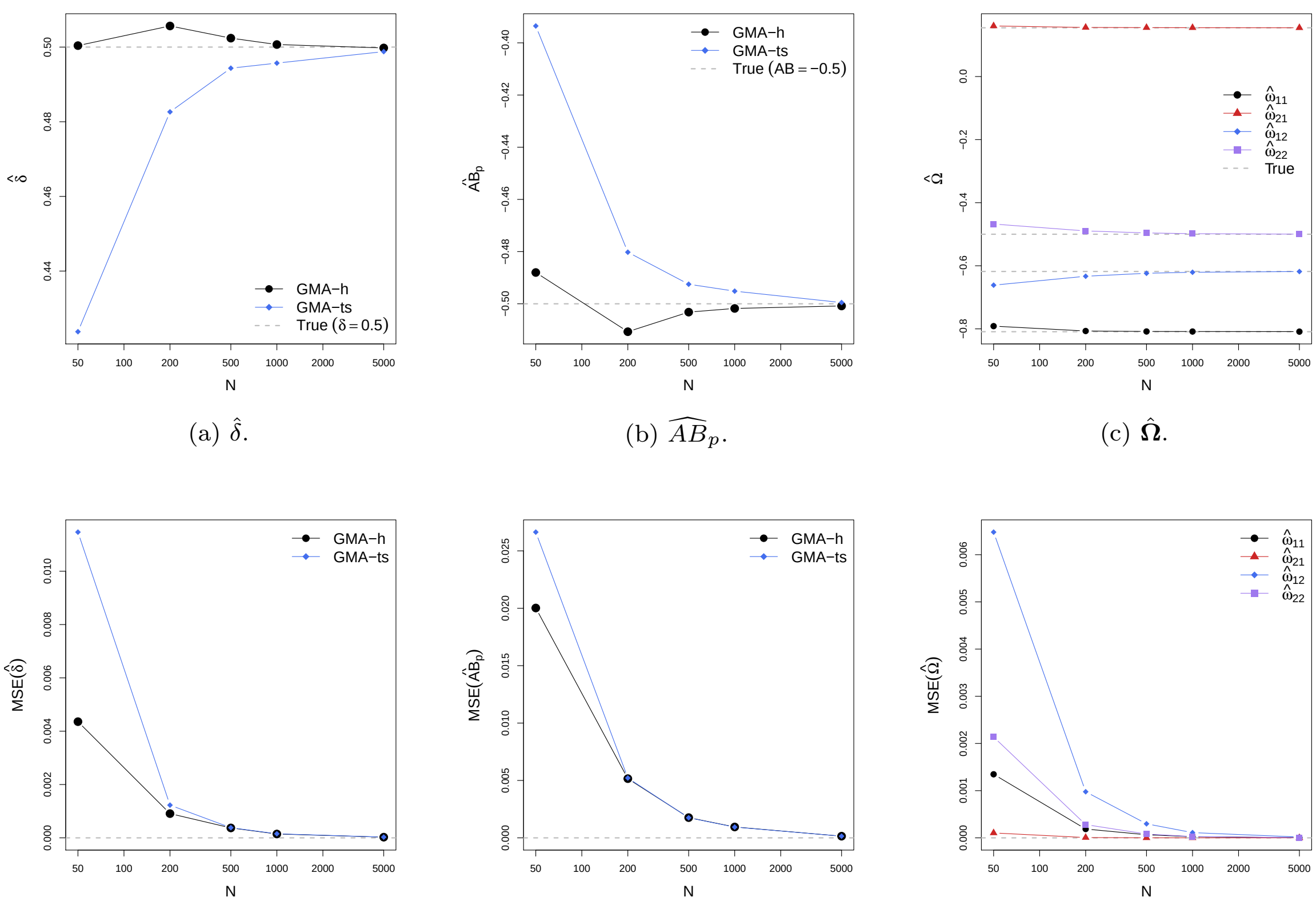

(a) $\hat{\delta}$. (b) $\widehat{AB}_p$. (c) $\hat{\mathbf{\Omega}}$.

(d) MSE of $\hat{\delta}$. (e) MSE of $\widehat{AB}_p$. (f) MSE of $\hat{\mathbf{\Omega}}$.

**Figure 7**: Average point estimates and MSEs of $\hat{\delta}$, $\widehat{AB}$ from GMA-h and GMA-ts, as well as $\hat{\mathbf{\Omega}}$ from GMA-h.

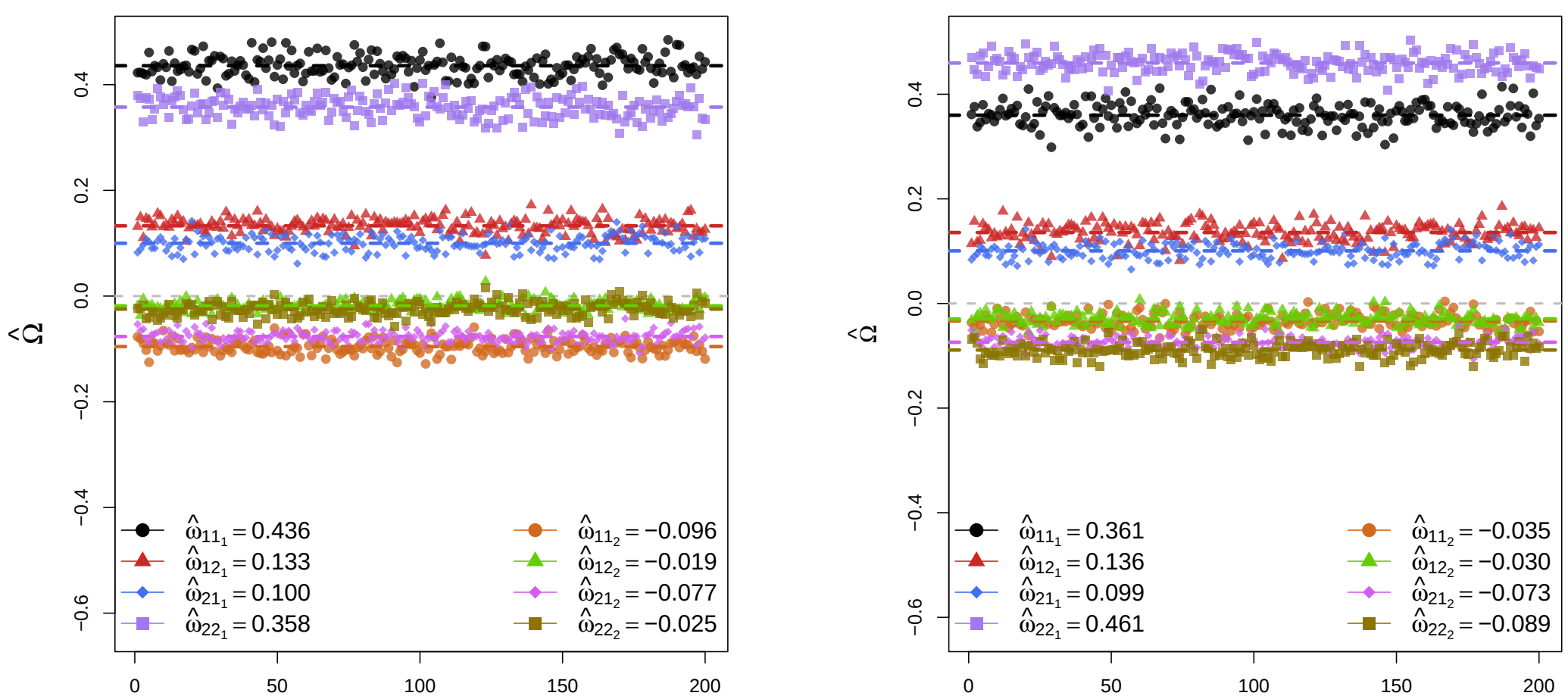

(a) GMA-h. (b) MAR.

**Figure 8**: Estimates of the transition matrix $\boldsymbol{\Omega}$ using (a) the GMA-h method and (b) the MAR method, from 200 bootstrap samples.

Table 1: Average point estimates under the single level models over 1000 runs. The values in the parathesis are empirical standard errors. The power is calculated using the asymptotic distributions for all the methods.

| $\delta$ | Method | $C$ | | $AB$ | | $\sigma_1^2$ | $\sigma_2^2$ | $\boldsymbol{\Omega}$ | | | |
|---|---|---|---|---|---|---|---|---|---|---|---|
| | | Estimate (SE) | Power | Estimate (SE) | Power | | | $\omega_{11}$ | $\omega_{12}$ | $\omega_{21}$ | $\omega_{22}$ |
| | True | 0.5 | | -0.5 | | 1 | 4 | -0.809 | -0.618 | 0.154 | -0.500 |
| | GMA-$\delta$ | 0.498 (0.355) | 0.418 | -0.496 (0.205) | 0.887 | 0.957 | 3.758 | -0.798 | -0.631 | 0.159 | -0.483 |
| 0.5 | GMA-0 | -0.000 (0.304) | 0.090 | 0.002 (0.091) | 0.019 | 0.957 | 2.818 | -0.640 | -0.473 | 0.159 | -0.641 |
| | MACC-$\delta$ | 0.499 (0.329) | 0.182 | -0.507 (0.210) | 0.693 | 1.982 | 7.727 | - | - | - | - |
| | BK | -0.008 (0.284) | 0.018 | -0.000 (0.113) | 0.027 | 1.982 | 5.795 | - | - | - | - |
| | True | 0.5 | | 0 ($B=0$) | | 1 | 4 | -0.809 | -0.618 | 0.154 | -0.500 |
| | GMA-$\delta$ | 0.498 (0.355) | 0.418 | 0.004 (0.106) | 0.041 | 0.957 | 3.758 | -0.798 | -0.631 | 0.159 | -0.483 |
| 0.5 | GMA-0 | -0.000 (0.304) | 0.090 | 0.502 (0.198) | 0.889 | 0.957 | 2.818 | -0.640 | -0.473 | 0.159 | -0.641 |
| | MACC-$\delta$ | 0.499 (0.329) | 0.182 | 0.002 (0.274) | 0.042 | 1.982 | 7.727 | - | - | - | - |
| | BK | -0.008 (0.284) | 0.018 | 0.510 (0.174) | 0.699 | 1.982 | 5.795 | - | - | - | - |
| | True | 0 | | -0.5 | | 1 | 4 | -0.809 | -0.618 | 0.154 | -0.500 |
| | GMA-$\delta$ | -0.002 (0.355) | 0.106 | -0.496 (0.205) | 0.887 | 0.957 | 3.758 | -0.798 | -0.631 | 0.159 | -0.483 |
| 0.5 | GMA-0 | -0.500 (0.304) | 0.510 | 0.002 (0.091) | 0.019 | 0.957 | 2.818 | -0.640 | -0.473 | 0.159 | -0.641 |
| | MACC-$\delta$ | -0.001 (0.328) | 0.016 | -0.507 (0.210) | 0.693 | 1.982 | 7.727 | - | - | - | - |
| | BK | -0.508 (0.284) | 0.265 | -0.000 (0.113) | 0.027 | 1.982 | 5.795 | - | - | - | - |
| | True | 0.5 | | -0.5 | | 1 | 4 | -0.500 | -1.000 | 0.250 | -0.500 |
| 0 | GMA-$\delta$ | 0.496 (0.296) | 0.401 | -0.496 (0.174) | 0.867 | 0.956 | 3.766 | -0.490 | -1.013 | 0.251 | -0.494 |
| | BK | 0.502 (0.410) | 0.188 | -0.509 (0.225) | 0.688 | 1.973 | 7.823 | - | - | - | - |

Table 2: Average estimates and 95% confidence intervals from GMA-h, GMA-$\delta$, MACC-h, KKB and BK for the fMRI data set using 200 bootstrap samples.

| Method | $\delta$ | $C$ | $AB$ |
|---|---|---|---|
| GMA-h | -0.370 (-0.530, -0.156) | -1.729 (-2.445, -0.964) | -0.623 (-1.239, 0.033) |
| GMA-ts | -0.343 (-0.501, -0.163) | -1.722 (-2.461, -0.904) | -0.604 (-1.234, 0.055) |
| MACC-h | -0.762 (-0.799, -0.721) | -2.310 (-2.958, -1.641) | -0.391 (-1.271, 0.465) |
| KKB | - | -2.513 (-2.922, -2.073) | -0.140 (-0.467, 0.196) |
| BK | - | -2.583 (-3.023, -2.142) | -0.146 (-0.467, 0.211) |

# Supplementary to Granger Mediation Analysis of Multiple Time Series with an Application to fMRI

Yi Zhao and Xi Luo

Department of Biostatistics, Brown University

## A Theory and Proof

In this section, we present the proof of theorems in the main text, as well as some additional results.

### A.1 Identifiability of causal estimands

We first proof the identifiability of causal estimands under imposed causal assumptions. Under the potential outcome framework, let $\{Z_t\}_t$ be the sequentially randomly assigned treatments, $\{(M_t(z_t, \{z_s\}_{s<t}), R_t(z_t, \{z_s\}_{s<t}))\}_t$ the bivariate potential mediator and outcome and $\{(M_t(Z_t, \{Z_s\}_{s<t}), R_t(Z_t, \{Z_s\}_{s<t}))\}_t$ the bivariate observed data. In Theorem 2, we show that the estimator of causal parameters are functions of correlation parameter $\delta$ and observed data. Here, we use a generic function $h_\delta(\cdot)$ to represent these estimators. Let $y_t = ((M_t(z_t, \{z_s\}_{s<t}), R_t(z_t, \{z_s\}_{s<t})), z_t)^\top$, $\boldsymbol{y} = \{y_t\}_t$ and $Y_t = ((M_t(Z_t, \{Z_s\}_{s<t}), R_t(Z_t, \{Z_s\}_{s<t})), Z_t)^\top$,

$\boldsymbol{Y} = \{Y_t\}_t$. We can show that

$$
\begin{aligned}
& \mathbb{E}\left[h_\delta(\{(M_t(z_t, \{z_s\}_{s<t}), R_t(z_t, \{z_s\}_{s<t})), z_t\})\right] \\
= & \int h_\delta(\boldsymbol{y}) f(\{(M_t(z_t, \{z_s\}_{s<t}), R_t(z_t, \{z_s\}_{s<t})), z_t\}) \, \mathrm{d}\boldsymbol{y} \\
= & \int h_\delta(\boldsymbol{y}) \prod_t f(((M_t(z_t, \{z_s\}_{s<t}), R_t(z_t, \{z_s\}_{s<t})), z_t) \mid \{Z_s = z_s\}_{s<t}) \, \mathrm{d}\boldsymbol{y} \\
= & \int h_\delta(\boldsymbol{y}) \prod_t f(((M_t(z_t, \{z_s\}_{s<t}), R_t(z_t, \{z_s\}_{s<t})), z_t) \mid Z_t = z_t, \{Z_s = z_s\}_{s<t}) \, \mathrm{d}\boldsymbol{y} \\
= & \int h_\delta(\boldsymbol{Y}) \prod_t f(((M_t(Z_t, \{Z_s\}_{s<t}), R_t(Z_t, \{Z_s\}_{s<t})), Z_t) \mid Z_t = z_t, \{Z_s = z_s\}_{s<t}) \, \mathrm{d}\boldsymbol{Y}.
\end{aligned}
$$

From line 3 to line 4, we use the sequential randomization assumption (A4); and from line 4 to line 5, use assumption (A3). Therefore, our estimators in Theorem 2 has causal interpretations.

## A.2 Proof of Lemma 1

*Proof.* For $\forall\ A, C, B$, $\boldsymbol{D}$ cannot be a zero matrix. Thus, we can multiply $\boldsymbol{D}^\top$ on both sides of Equation (7), and yield

$$\left(\boldsymbol{D}^\top \boldsymbol{D}\right) \boldsymbol{\omega}_j = \boldsymbol{D}^\top \boldsymbol{\eta}_j.$$

If $\boldsymbol{D}^\top \boldsymbol{D}$ is of full rank, then we prove the lemma.

$$
\boldsymbol{D}^\top \boldsymbol{D} = \begin{pmatrix}
A^2+1 & AC-B & 0 & 0 \\
AC-B & C^2+B^2+1 & 0 & 0 \\
0 & 0 & A^2+1 & AC-B \\
0 & 0 & AC-B & C^2+B^2+1
\end{pmatrix},
$$

which is a block matrix and is of full rank for $\forall\ A, C, B \in \mathbb{R}$, since

$$\det\left(\begin{pmatrix} A^2+1 & AC-B \\ AC-B & C^2+B^2+1 \end{pmatrix}\right) = (AB+C)^2 + A^2 + 1 > 0, \quad \forall\ A, C, B \in \mathbb{R}.$$

Therefore, $\boldsymbol{\omega}_j$ is uniquely determined as

$$\boldsymbol{\omega}_j = \left(\boldsymbol{D}^\top \boldsymbol{D}\right)^{-1} \boldsymbol{D}^\top \boldsymbol{\eta}_j.$$

□

## A.3 Proof of Theorem 2

*Proof.* Under models (5) and (6), at time $t$ ($t \geq p+1$), the conditional distribution of $M_t$ and $R_t$ are

$$\begin{aligned} M_t \mid \boldsymbol{X}_t &\sim \mathcal{N}\left(\mu_{(M)t}, \sigma_1^2\right), \\ R_t \mid M_t, \boldsymbol{X}_t &\sim \mathcal{N}\left(\mu_{(R|M)t}, \sigma_2^2(1-\delta^2)\right), \end{aligned}$$

where

$$\begin{aligned} \mu_{(M)t} &= Z_t A + \boldsymbol{\phi}_1^\top \boldsymbol{Z}_t^{(p)} + \boldsymbol{\psi}_{11}^\top \boldsymbol{M}_t^{(p)} + \boldsymbol{\psi}_{21}^\top \boldsymbol{R}_t^{(p)} = \boldsymbol{X}_t^\top \boldsymbol{\theta}_1, \\ \mu_{(R|M)t} &= Z_t(C-\kappa A) + M_t(B+\kappa) + (\boldsymbol{\phi}_2 - \kappa\boldsymbol{\phi}_1)^\top \boldsymbol{Z}_t^{(p)} + (\boldsymbol{\psi}_{12} - \kappa\boldsymbol{\psi}_{11})^\top \boldsymbol{M}_t^{(p)} + (\boldsymbol{\psi}_{22} - \kappa\boldsymbol{\psi}_{21})^\top \boldsymbol{R}_t^{(p)} \\ &= M_t B + \boldsymbol{X}_t^\top \boldsymbol{\theta}_2 + \kappa(M_t - \boldsymbol{X}_t^\top \boldsymbol{\theta}_1), \end{aligned}$$

and $\kappa = \delta\sigma_2/\sigma_1$.

Given the initial $p$ observations, the joint density function of $\{M_t, R_t\}_{t=p+1}^{T}$ is

$$f\left((M_{p+1}, R_{p+1}), \ldots, (M_T, R_T)\right) = \prod_{t=p+1}^{T} f\left(R_t \mid M_t, \boldsymbol{X}_t\right) f\left(M_t \mid \boldsymbol{X}_t\right),$$

and the conditional log-likelihood is

$$\begin{aligned}
&\ell\left(\boldsymbol{\theta}_1, \boldsymbol{\theta}_2, B, \boldsymbol{\Sigma}\right) \\
&= \sum_{t=p+1}^{T} \left[\log f\left(R_t \mid M_t, \boldsymbol{X}_t\right) + \log f\left(M_t \mid \boldsymbol{X}_t\right)\right] \\
&= -\frac{T-p}{2} \log \sigma_1^2 \sigma_2^2 (1-\delta^2) - \sum_{t=p+1}^{T} \left[\frac{1}{2\sigma_1^2}(M_t - \mu_{(M)t})^2 + \frac{1}{2\sigma_2^2(1-\delta^2)}(R_t - \mu_{(R|M)t})^2\right] \\
&= -\frac{T-p}{2} \log \sigma_1^2 \sigma_2^2 (1-\delta^2) - \frac{1}{2\sigma_1^2}\left(\boldsymbol{M} - \boldsymbol{X}\boldsymbol{\theta}_1\right)^\top \left(\boldsymbol{M} - \boldsymbol{X}\boldsymbol{\theta}_1\right) \\
&\quad - \frac{1}{2\sigma_2^2(1-\delta^2)} \left((\boldsymbol{R} - \boldsymbol{M}B - \boldsymbol{X}\boldsymbol{\theta}_2) - \kappa(\boldsymbol{M} - \boldsymbol{X}\boldsymbol{\theta}_1)\right)^\top \left((\boldsymbol{R} - \boldsymbol{M}B - \boldsymbol{X}\boldsymbol{\theta}_2) - \kappa(\boldsymbol{M} - \boldsymbol{X}\boldsymbol{\theta}_1)\right),
\end{aligned}$$

where $\boldsymbol{R} = (R_{p+1}, \ldots, R_T)^\top$, $\boldsymbol{M} = (M_{p+1}, \ldots, M_T)^\top$, $\boldsymbol{X} = (\boldsymbol{X}_{p+1}, \ldots, \boldsymbol{X}_T)^\top$. To maximize $\ell$, we fix the $\delta$ value, and take partial derivatives over the rest parameters. First, we start with taking partial derivative over $B$,

$$\frac{\partial \ell}{\partial B} = \frac{1}{\sigma_2^2(1-\delta^2)} \left[-\boldsymbol{M}^\top \boldsymbol{M} B + (\boldsymbol{M}^\top (\boldsymbol{R} - \boldsymbol{X}\boldsymbol{\theta}_2) - \boldsymbol{R}\boldsymbol{M}^\top (\boldsymbol{M} - \boldsymbol{X}\boldsymbol{\theta}_1))\right] = 0,$$

$$\Rightarrow \quad \hat{B} = (\boldsymbol{M}^\top \boldsymbol{M})^{-1} \boldsymbol{M}^\top \left((\boldsymbol{R} - \boldsymbol{X}\boldsymbol{\theta}_2) - \kappa(\boldsymbol{M} - \boldsymbol{X}\boldsymbol{\theta}_1)\right).$$

We plug $\hat{B}$ into $\ell$, and yield

$$\begin{aligned}\ell_{\boldsymbol{\theta}_2}(\hat{B}) = &-\frac{1}{2\sigma_2^2(1-\delta^2)}\left((\mathbf{I}-\boldsymbol{P}_M)\boldsymbol{R}-(\mathbf{I}-\boldsymbol{P}_M)\boldsymbol{X}\boldsymbol{\theta}_2+\kappa\boldsymbol{P}_M(\boldsymbol{M}-\boldsymbol{X}\boldsymbol{\theta}_1)\right)^\top \\ &\quad\left((\mathbf{I}-\boldsymbol{P}_M)\boldsymbol{R}-(\mathbf{I}-\boldsymbol{P}_M)\boldsymbol{X}\boldsymbol{\theta}_2+\kappa\boldsymbol{P}_M(\boldsymbol{M}-\boldsymbol{X}\boldsymbol{\theta}_1)\right) \\ &+\frac{\kappa}{\sigma_2^2(1-\delta^2)}\left((\mathbf{I}-\boldsymbol{P}_M)\boldsymbol{R}-(\mathbf{I}-\boldsymbol{P}_M)\boldsymbol{X}\boldsymbol{\theta}_2+\kappa\boldsymbol{P}_M(\boldsymbol{M}-\boldsymbol{X}\boldsymbol{\theta}_1)\right)^\top(\boldsymbol{M}-\boldsymbol{X}\boldsymbol{\theta}_1),\end{aligned}$$

where $\boldsymbol{P}_M = \boldsymbol{M}(\boldsymbol{M}^\top\boldsymbol{M})^{-1}\boldsymbol{M}^\top$ is the projection matrix of $\boldsymbol{M}$. Next, we optimize the function over $\boldsymbol{\theta}_2$,

$$\frac{\partial \ell_{\boldsymbol{\theta}_2}(\hat{B})}{\partial \boldsymbol{\theta}_2} = -\frac{1}{\sigma_2^2(1-\delta^2)}\left[\boldsymbol{X}^\top(\mathbf{I}-\boldsymbol{P}_M)\boldsymbol{X}\boldsymbol{\theta}_2-\left(\boldsymbol{X}^\top(\mathbf{I}-\boldsymbol{P}_M)\boldsymbol{R}-\kappa\boldsymbol{X}^\top(\mathbf{I}-\boldsymbol{P}_M)(\boldsymbol{M}-\boldsymbol{X}\boldsymbol{\theta}_1)\right)\right]=0,$$

$$\Rightarrow\quad \hat{\boldsymbol{\theta}}_2 = \left(\boldsymbol{X}^\top(\mathbf{I}-\boldsymbol{P}_M)\boldsymbol{X}\right)^{-1}\boldsymbol{X}^\top(\mathbf{I}-\boldsymbol{P}_M)(\boldsymbol{R}+\kappa\boldsymbol{X}\boldsymbol{\theta}_1),$$

and this yields the profile function of $\boldsymbol{\theta}_1$ as

$$\begin{aligned}\ell_{\boldsymbol{\theta}_1}(\hat{B},\hat{\boldsymbol{\theta}}_2) = &-\frac{1}{2\sigma_1^2}(\boldsymbol{M}-\boldsymbol{X}\boldsymbol{\theta}_1)^\top(\boldsymbol{M}-\boldsymbol{X}\boldsymbol{\theta}_1)-\frac{\kappa^2}{2\sigma_2^2(1-\delta^2)}(\boldsymbol{M}-\boldsymbol{X}\boldsymbol{\theta}_1)^\top(\boldsymbol{M}-\boldsymbol{X}\boldsymbol{\theta}_1) \\ &-\frac{1}{2\sigma_2^2(1-\delta^2)}\left((\mathbf{I}-\boldsymbol{P}_M-\boldsymbol{P}_{MX})\boldsymbol{R}+\kappa\boldsymbol{M}-\kappa(\boldsymbol{P}_{MX}+\boldsymbol{P}_M)\boldsymbol{X}\boldsymbol{\theta}_1\right)^\top \\ &\quad\left((\mathbf{I}-\boldsymbol{P}_M-\boldsymbol{P}_{MX})\boldsymbol{R}+\kappa\boldsymbol{M}-\kappa(\boldsymbol{P}_{MX}+\boldsymbol{P}_M)\boldsymbol{X}\boldsymbol{\theta}_1\right) \\ &+\frac{\kappa}{\sigma_2^2(1-\delta^2)}\left((\mathbf{I}-\boldsymbol{P}_M-\boldsymbol{P}_{MX})\boldsymbol{R}+\kappa\boldsymbol{M}\right)^\top(\boldsymbol{M}-\boldsymbol{X}\boldsymbol{\theta}_1) \\ &-\frac{\kappa^2}{\sigma_2^2(1-\delta^2)}\boldsymbol{\theta}_1^\top\boldsymbol{X}^\top(\boldsymbol{P}_{MX}+\boldsymbol{P}_M)(\boldsymbol{M}-\boldsymbol{X}\boldsymbol{\theta}_1),\end{aligned}$$

where $\boldsymbol{P}_{MX} = (\mathbf{I}-\boldsymbol{P}_M)\boldsymbol{X}(\boldsymbol{X}^\top(\mathbf{I}-\boldsymbol{P}_M)\boldsymbol{X})^{-1}\boldsymbol{X}^\top(\mathbf{I}-\boldsymbol{P}_M)$. Lastly, using this profile log-likelihood function, we have

$$\frac{\partial \ell_{\boldsymbol{\theta}_1}(\hat{B},\hat{\boldsymbol{\theta}}_2)}{\partial \boldsymbol{\theta}_1} = -\frac{1}{\sigma_1^2}\boldsymbol{X}^\top\boldsymbol{X}\boldsymbol{\theta}_1+\frac{1}{\sigma_1^2}\boldsymbol{X}^\top\boldsymbol{M}=0,$$

$$\Rightarrow \quad \hat{\boldsymbol{\theta}}_1 = (\boldsymbol{X}^\top \boldsymbol{X})^{-1} \boldsymbol{X}^\top \boldsymbol{M},$$

which is independent of $\delta$. To estimate the variances, by plugging in the maximizers of $(\boldsymbol{\theta}_1, \boldsymbol{\theta}_2, B)$, we have

$$\ell(\hat{\boldsymbol{\theta}}_1, \hat{\boldsymbol{\theta}}_2, \hat{B}) = -\frac{T-p}{2} \log \sigma_1^2 \sigma_2^2 (1-\delta^2) - \frac{1}{2\sigma_1^2} \boldsymbol{M}^\top (\mathbf{I} - \boldsymbol{P}_X) \boldsymbol{M} - \frac{1}{2\sigma_2^2(1-\delta^2)} \boldsymbol{R}^\top (\mathbf{I} - \boldsymbol{P}_{MX} - \boldsymbol{P}_M) \boldsymbol{R}.$$

Therefore,

$$\frac{\ell(\hat{\boldsymbol{\theta}}_1, \hat{\boldsymbol{\theta}}_2, \hat{B})}{\partial \sigma_1^2} = -\frac{T-p}{2\sigma_1^2} + \frac{1}{2\sigma_1^4} \boldsymbol{M}^\top (\mathbf{I} - \boldsymbol{P}_X) \boldsymbol{M} = 0, \quad \Rightarrow \quad \hat{\sigma}_1^2 = \frac{1}{T-p} \boldsymbol{M}^\top (\mathbf{I} - \boldsymbol{P}_X) \boldsymbol{M},$$

$$\frac{\ell(\hat{\boldsymbol{\theta}}_1, \hat{\boldsymbol{\theta}}_2, \hat{B})}{\partial \sigma_2^2} = -\frac{T-p}{2\sigma_2^2} + \frac{1}{2\sigma_2^2(1-\delta^2)} \boldsymbol{R}^\top (\mathbf{I} - \boldsymbol{P}_{MX} - \boldsymbol{P}_M) \boldsymbol{R},$$

$$\Rightarrow \quad \hat{\sigma}_2^2 = \frac{1}{(T-p)(1-\delta^2)} \boldsymbol{R}^\top (\mathbf{I} - \boldsymbol{P}_{MX} - \boldsymbol{P}_M) \boldsymbol{R},$$

where $\boldsymbol{P}_X = \boldsymbol{X}(\boldsymbol{X}^\top \boldsymbol{X})^{-1} \boldsymbol{X}^\top$. Then, we replace the unknown parameters in the formulas with its corresponding maximizers, and this finishes the proof. □

## A.4 Asymptotic properties of the CMLE in Theorem 2

**Theorem A.1.** *Let $\boldsymbol{\theta}^\top = \left(\boldsymbol{\theta}_1^\top, \boldsymbol{\theta}_2^\top, B\right)$, under the stationary condition, the CMLE of $\boldsymbol{\theta}$ in Theorem 2 are consistent with*

$$\sqrt{T}\left(\hat{\boldsymbol{\theta}} - \boldsymbol{\theta}\right) \xrightarrow{\mathcal{D}} \mathcal{N}\left(\mathbf{0}, I^{-1}(\boldsymbol{\theta})\right), \tag{1}$$

*where*

$$I(\boldsymbol{\theta}) = \mathbb{E}\left[-\nabla^2 \ell(\boldsymbol{\theta})\right] \tag{2}$$

*is the Fisher information matrix.*

*Proof.* Following the second Bartlett identity, we first calculate the second order partial derivatives of $\boldsymbol{\theta}$. Given the conditional log-likelihood function (10), we have the following first order partial derivatives,

$$
\begin{aligned}
\frac{\partial \ell}{\partial \boldsymbol{\theta}_1} &= \frac{1}{\sigma_1^2(1-\delta^2)}\boldsymbol{X}^\top(\boldsymbol{M}-\boldsymbol{X}\boldsymbol{\theta}) = \frac{\kappa}{\sigma_2^2(1-\delta^2)}\boldsymbol{X}^\top(\boldsymbol{R}-\boldsymbol{M}B-\boldsymbol{X}\boldsymbol{\theta}_2),\\
\frac{\partial \ell}{\partial \boldsymbol{\theta}_2} &= \frac{1}{\sigma_2^2(1-\delta^2)}\boldsymbol{X}^\top(\boldsymbol{R}-\boldsymbol{M}B-\boldsymbol{X}\boldsymbol{\theta}_2) - \frac{\kappa}{\sigma_2^2(1-\delta^2)}\boldsymbol{X}^\top(\boldsymbol{M}-\boldsymbol{X}\boldsymbol{\theta}_1),\\
\frac{\partial \ell}{\partial B} &= \frac{1}{\sigma_2^2(1-\delta^2)}\boldsymbol{M}^\top(\boldsymbol{R}-\boldsymbol{M}B-\boldsymbol{X}\boldsymbol{\theta}_2) - \frac{\kappa}{\sigma_2^2(1-\delta^2)}\boldsymbol{M}^\top(\boldsymbol{M}-\boldsymbol{X}\boldsymbol{\theta}_1);
\end{aligned}
$$

and the second order

$$
\begin{aligned}
\frac{\partial^2 \ell}{\partial \boldsymbol{\theta}_1 \partial \boldsymbol{\theta}_1^\top} &= -\frac{1}{\sigma_1^2(1-\delta^2)}\boldsymbol{X}^\top\boldsymbol{X}, \quad \frac{\partial^2 \ell}{\partial \boldsymbol{\theta}_1 \partial \boldsymbol{\theta}_2^\top} = \frac{\kappa}{\sigma_2^2(1-\delta^2)}\boldsymbol{X}^\top\boldsymbol{X}, \quad \frac{\partial^2 \ell}{\partial \boldsymbol{\theta}_1 \partial B} = \frac{\kappa}{\sigma_2^2(1-\delta^2)}\boldsymbol{X}^\top\boldsymbol{M},\\
\frac{\partial^2 \ell}{\partial \boldsymbol{\theta}_2 \partial \boldsymbol{\theta}_2^\top} &= -\frac{1}{\sigma_2^2(1-\delta^2)}\boldsymbol{X}^\top\boldsymbol{X}, \quad \frac{\partial^2 \ell}{\partial \boldsymbol{\theta}_2 \partial B} = -\frac{1}{\sigma_2^2(1-\delta^2)}\boldsymbol{X}^\top\boldsymbol{M}, \quad \frac{\partial^2 \ell}{\partial B^2} = -\frac{1}{\sigma_2^2(1-\delta^2)}\boldsymbol{M}^\top\boldsymbol{M}.
\end{aligned}
$$

Thus,

$$
\nabla^2 \ell(\boldsymbol{\theta}) = \begin{pmatrix}
-\frac{1}{\sigma_1^2(1-\delta^2)}\boldsymbol{X}^\top\boldsymbol{X} & \frac{\kappa}{\sigma_2^2(1-\delta^2)}\boldsymbol{X}^\top\boldsymbol{X} & \frac{\kappa}{\sigma_2^2(1-\delta^2)}\boldsymbol{X}^\top\boldsymbol{M}\\
 & -\frac{1}{\sigma_2^2(1-\delta^2)}\boldsymbol{X}^\top\boldsymbol{X} & -\frac{1}{\sigma_2^2(1-\delta^2)}\boldsymbol{X}^\top\boldsymbol{M}\\
 & & -\frac{1}{\sigma_2^2(1-\delta^2)}\boldsymbol{M}^\top\boldsymbol{M}
\end{pmatrix},
$$

where

$$
\boldsymbol{X}^\top\boldsymbol{X} = \sum_{t=p+1}^{T}\boldsymbol{X}_t\boldsymbol{X}_t^\top, \quad \boldsymbol{X}^\top\boldsymbol{M} = \sum_{t=p+1}^{T}\boldsymbol{X}_t M_t, \quad \boldsymbol{M}^\top\boldsymbol{M} = \sum_{t=p+1}^{T} M_t^2.
$$

Now we first calculate $\mathbb{E}\left[\boldsymbol{X}_t^\top \boldsymbol{X}_t\right]$.

$$\boldsymbol{X}_t\boldsymbol{X}_t^\top = \begin{pmatrix} Z_t^2 & Z_t\boldsymbol{Z}_{t-1}^{(p)\top} & Z_t\boldsymbol{M}_{t-1}^{(p)\top} & Z_t\boldsymbol{R}_{t-1}^{(p)\top} \\ & \boldsymbol{Z}_{t-1}^{(p)}\boldsymbol{Z}_{t-1}^{(p)\top} & \boldsymbol{Z}_{t-1}^{(p)}\boldsymbol{M}_{t-1}^{(p)\top} & \boldsymbol{Z}_{t-1}^{(p)}\boldsymbol{R}_{t-1}^{(p)\top} \\ & & \boldsymbol{M}_{t-1}^{(p)}\boldsymbol{M}_{t-1}^{(p)\top} & \boldsymbol{M}_{t-1}^{(p)}\boldsymbol{R}_{t-1}^{(p)\top} \\ & & & \boldsymbol{R}_{t-1}^{(p)}\boldsymbol{R}_{t-1}^{(p)\top} \end{pmatrix}.$$

From our proposed models (1) and (2),

$$\begin{cases} M_t = Z_tA + E_{1t} \\ R_t = Z_tC + M_tB + E_{2t} \end{cases},$$

we have the same formulation for the vector representations as

$$\begin{cases} \boldsymbol{M}_{t-1}^{(p)\top} = \boldsymbol{Z}_{t-1}^{(p)\top}A + \boldsymbol{E}_{1,t-1}^{(p)\top} \\ \boldsymbol{R}_{t-1}^{(p)\top} = \boldsymbol{Z}_{t-1}^{(p)\top}C + \boldsymbol{M}_{t-1}^{(p)\top}B + \boldsymbol{E}_{2,t-1}^{(p)\top} \end{cases}. \tag{3}$$

Let $\boldsymbol{E}_t = (E_{1t}, E_{2t})^\top$, $\boldsymbol{\epsilon}_t = (\epsilon_{1t}, \epsilon_{2t})^\top$, we have the matrix form of the MAR($p$) model (3) as

$$\boldsymbol{E}_t = \sum_{j=1}^{p} \boldsymbol{\Omega}_j^\top \boldsymbol{E}_{t-j} + \boldsymbol{\epsilon}_t.$$

From the model, the current $\boldsymbol{E}_t$ depends on $(\boldsymbol{E}_{t-p}, \dots, \boldsymbol{E}_{t-1})$. To derive the covariance matrix, we let

$$\boldsymbol{\xi}_t = \begin{pmatrix} \boldsymbol{E}_t \\ \boldsymbol{E}_{t-1} \\ \vdots \\ \boldsymbol{E}_{t-p+1} \end{pmatrix}, \quad \boldsymbol{v}_t = \begin{pmatrix} \boldsymbol{\epsilon}_t \\ \mathbf{0} \\ \vdots \\ \mathbf{0} \end{pmatrix},$$

then

$$\boldsymbol{\xi}_t = \boldsymbol{F}\boldsymbol{\xi}_{t-1} + \boldsymbol{v}_t,$$

where $\boldsymbol{F}$ is the companion matrix, and

$$\mathrm{Cov}(\boldsymbol{v}_t) = \begin{pmatrix} \boldsymbol{\Sigma} & \mathbf{0} & \cdots & \mathbf{0} \\ \mathbf{0} & \mathbf{0} & \cdots & \mathbf{0} \\ \vdots & \vdots & \ddots & \vdots \\ \mathbf{0} & \mathbf{0} & \cdots & \mathbf{0} \end{pmatrix} \triangleq \boldsymbol{\Xi}.$$

Therefore,

$$\boldsymbol{\Pi} \triangleq \mathrm{Cov}(\boldsymbol{\xi}_t) = \boldsymbol{F}\mathrm{Cov}(\boldsymbol{\xi}_{t-1})\boldsymbol{F}^\top + \mathrm{Cov}(\boldsymbol{v}_t) = \boldsymbol{F}\boldsymbol{\Pi}\boldsymbol{F}^\top + \boldsymbol{\Xi}.$$

Taking the vectorization on both sides of the equation, we have

$$\mathrm{vec}(\boldsymbol{\Pi}) = (\boldsymbol{F} \otimes \boldsymbol{F})\,\mathrm{vec}(\boldsymbol{\Pi}) + \mathrm{vec}(\boldsymbol{\Xi}), \quad \Rightarrow \quad \mathrm{vec}(\boldsymbol{\Pi}) = (\mathbf{I} - \boldsymbol{F} \otimes \boldsymbol{F})^{-1}\,\mathrm{vec}(\boldsymbol{\Xi}),$$

$$\boldsymbol{\Pi} = \mathrm{Cov}(\boldsymbol{\xi}_t) = \mathbb{E}(\boldsymbol{\xi}_t\boldsymbol{\xi}_t^\top) = \begin{pmatrix} \boldsymbol{\Gamma}_0 & \boldsymbol{\Gamma}_1 & \cdots & \boldsymbol{\Gamma}_{p-1} \\ \boldsymbol{\Gamma}_1^\top & \boldsymbol{\Gamma}_0 & \cdots & \boldsymbol{\Gamma}_{p-2} \\ \vdots & \vdots & \ddots & \vdots \\ \boldsymbol{\Gamma}_{p-1}^\top & \boldsymbol{\Gamma}_{p-2}^\top & \cdots & \boldsymbol{\Gamma}_0 \end{pmatrix}, \quad \text{where } \begin{cases} \boldsymbol{\Gamma}_0 = \mathbb{E}\left(\boldsymbol{E}_t\boldsymbol{E}_t^\top\right) \\ \boldsymbol{\Gamma}_j = \mathbb{E}\left(\boldsymbol{E}_t\boldsymbol{E}_{t-j}^\top\right) \end{cases}.$$

For $1 \leq j \leq p$, the covariance between $\boldsymbol{\xi}_t$ and $\boldsymbol{\xi}_{t-j}$ ($j$-step time lag) is

$$\boldsymbol{\Pi}_j \triangleq \mathbb{E}(\boldsymbol{\xi}_t\boldsymbol{\xi}_{t-j}^\top) = \boldsymbol{F}\mathbb{E}(\boldsymbol{\xi}_{t-1}\boldsymbol{\xi}_{t-j}^\top) + \mathbb{E}(\boldsymbol{v}_t\boldsymbol{\xi}_{t-j}^\top) = \boldsymbol{F}\boldsymbol{\Pi}_{j-1} = \cdots = \boldsymbol{F}^j\boldsymbol{\Pi}.$$

Let $\boldsymbol{e}_1^{(2)} = (1,0)^\top$ and $\boldsymbol{e}_2^{(2)} = (0,1)^\top$, then

$$\begin{cases} \boldsymbol{E}_{1,t-1}^{(p)} = (\mathbf{I}_p \otimes \boldsymbol{e}_1^{(2)\top})\boldsymbol{\xi}_{t-1} \triangleq \boldsymbol{J}_1^\top \boldsymbol{\xi}_{t-1} \\ \boldsymbol{E}_{2,t-1}^{(p)} = (\mathbf{I}_p \otimes \boldsymbol{e}_2^{(2)\top})\boldsymbol{\xi}_{t-1} \triangleq \boldsymbol{J}_2^\top \boldsymbol{\xi}_{t-1} \end{cases},$$

and model (3) is represented as

$$\begin{cases} \boldsymbol{M}_{t-1}^{(p)} = \boldsymbol{Z}_{t-1}^{(p)} A + \boldsymbol{J}_1^\top \boldsymbol{\xi}_{t-1} \\ \boldsymbol{R}_{t-1}^{(p)} = \boldsymbol{Z}_{t-1}^{(p)} C + \boldsymbol{M}_{t-1}^{(p)} B + \boldsymbol{J}_2^\top \boldsymbol{\xi}_{t-1} = \boldsymbol{Z}_{t-1}^{(p)}(C + AB) + B\boldsymbol{J}_1^\top \boldsymbol{\xi}_{t-1} + \boldsymbol{J}_2^\top \boldsymbol{\xi}_{t-1} \end{cases}.$$

Assume $\mathbb{E}(Z_t) = 0$ (without loss of generality), $\mathbb{E}(Z_t^2) = q$, and under the randomization assumption, $Z_t \perp\!\!\!\perp Z_s$ for $t \neq s$ and $Z_t \perp\!\!\!\perp \boldsymbol{\xi}_s$ for $\forall\, t, s$, then we have

$$\begin{aligned}
\mathbb{E}\left(Z_t \boldsymbol{Z}_{t-1}^{(p)\top}\right) &= \mathbf{0}, \\
\mathbb{E}\left(Z_t \boldsymbol{M}_{t-1}^{(p)\top}\right) &= \mathbb{E}\left(Z_t \left(\boldsymbol{Z}_{t-1}^{(p)} A + \boldsymbol{J}_1^\top \boldsymbol{\xi}_{t-1}\right)^\top\right) = \mathbf{0}, \\
\mathrm{E}\left(Z_t \boldsymbol{R}_{t-1}^{(p)\top}\right) &= \mathbb{E}\left(Z_t \left(\boldsymbol{Z}_{t-1}^{(p)}(C + AB) + B\boldsymbol{J}_1^\top \boldsymbol{\xi}_{t-1} + \boldsymbol{J}_2^\top \boldsymbol{\xi}_{t-1}\right)^\top\right) = \mathbf{0}, \\
\mathbb{E}\left(\boldsymbol{Z}_{t-1}^{(p)} \boldsymbol{Z}_{t-1}^{(p)\top}\right) &= q\mathbf{I}_p, \\
\mathbb{E}\left(\boldsymbol{Z}_{t-1}^{(p)} \boldsymbol{M}_{t-1}^{(p)\top}\right) &= \mathbb{E}\left(\boldsymbol{Z}_{t-1}^{(p)} \left(\boldsymbol{Z}_{t-1}^{(p)} A + \boldsymbol{J}_1^\top \boldsymbol{\xi}_{t-1}\right)^\top\right) = Aq\mathbf{I}_p, \\
\mathbb{E}\left(\boldsymbol{Z}_{t-1}^{(p)} \boldsymbol{R}_{t-1}^{(p)\top}\right) &= \mathbb{E}\left(\boldsymbol{Z}_{t-1}^{(p)} \left(\boldsymbol{Z}_{t-1}^{(p)}(C + AB) + B\boldsymbol{J}_1^\top \boldsymbol{\xi}_{t-1} + \boldsymbol{J}_2^\top \boldsymbol{\xi}_{t-1}\right)^\top\right) = (C + AB)q\mathbf{I}_p, \\
\mathbb{E}\left(\boldsymbol{M}_{t-1}^{(p)} \boldsymbol{M}_{t-1}^{(p)\top}\right) &= \mathbb{E}\left(\boldsymbol{Z}_{t-1}^{(p)} A + \boldsymbol{J}_1^\top \boldsymbol{\xi}_{t-1}\right)\left(\boldsymbol{Z}_{t-1}^{(p)} A + \boldsymbol{J}_1^\top \boldsymbol{\xi}_{t-1}\right)^\top = A^2 q\mathbf{I}_q + \boldsymbol{J}_1^\top \boldsymbol{\Pi} \boldsymbol{J}_1 \triangleq \boldsymbol{Q}_{\boldsymbol{MM}}, \\
\mathbb{E}\left(\boldsymbol{M}_{t-1}^{(p)} \boldsymbol{R}_{t-1}^{(p)\top}\right) &= \mathbb{E}\left(\boldsymbol{Z}_{t-1}^{(p)} A + \boldsymbol{J}_1^\top \boldsymbol{\xi}_{t-1}\right)\left(\boldsymbol{Z}_{t-1}^{(p)}(C + AB) + B\boldsymbol{J}_1^\top \boldsymbol{\xi}_{t-1} + \boldsymbol{J}_2^\top \boldsymbol{\xi}_{t-1}\right)^\top \\
&= A(C + AB)q\mathbf{I}_p + B\boldsymbol{J}_1^\top \boldsymbol{\Pi} \boldsymbol{J}_1 + \boldsymbol{J}_1^\top \boldsymbol{\Pi} \boldsymbol{J}_2 \triangleq \boldsymbol{Q}_{\boldsymbol{MR}}, \\
\mathbb{E}\left(\boldsymbol{R}_{t-1}^{(p)} \boldsymbol{R}_{t-1}^{(p)\top}\right) &= \mathbb{E}\left(\boldsymbol{Z}_{t-1}^{(p)}(C + AB) + B\boldsymbol{J}_1^\top \boldsymbol{\xi}_{t-1} + \boldsymbol{J}_2^\top \boldsymbol{\xi}_{t-1}\right)\left(\boldsymbol{Z}_{t-1}^{(p)}(C + AB) + B\boldsymbol{J}_1^\top \boldsymbol{\xi}_{t-1} + \boldsymbol{J}_2^\top \boldsymbol{\xi}_{t-1}\right)^\top \\
&= (C + AB)^2 q\mathbf{I}_p + B^2 \boldsymbol{J}_1^\top \boldsymbol{\Pi} \boldsymbol{J}_1 + B\boldsymbol{J}_1^\top \boldsymbol{\Pi} \boldsymbol{J}_2 + B\boldsymbol{J}_2^\top \boldsymbol{\Pi} \boldsymbol{J}_1 + \boldsymbol{J}_1^\top \boldsymbol{\Pi} \boldsymbol{J}_1 \triangleq \boldsymbol{Q}_{\boldsymbol{RR}}.
\end{aligned}$$

Using these quantities, we have $\mathbb{E}\left[\boldsymbol{X}_t^\top \boldsymbol{X}_t\right]$.

The next step is to calculate $\mathbb{E}\left[\boldsymbol{X}_t M_t\right]$, where

$$\boldsymbol{X}_t M_t = \begin{pmatrix} Z_t M_t \\ \boldsymbol{Z}_{t-1}^{(p)} M_t \\ \boldsymbol{M}_{t-1}^{(p)} M_t \\ \boldsymbol{R}_{t-1}^{(p)} M_t \end{pmatrix},$$

and

$$M_t = Z_t A + \boldsymbol{Z}_{t-1}^{(p)\top}\boldsymbol{\phi}_1 + \boldsymbol{M}_{t-1}^{(p)\top}\boldsymbol{\psi}_{11} + \boldsymbol{R}_{t-1}^{(p)\top}\boldsymbol{\psi}_{21} + \epsilon_{1t}.$$

We have

$$\begin{aligned}
\mathbb{E}\left(Z_t M_t\right) &= \mathbb{E}\left(Z_t\left(Z_t A + \boldsymbol{Z}_{t-1}^{(p)\top}\boldsymbol{\phi}_1 + \boldsymbol{M}_{t-1}^{(p)\top}\boldsymbol{\psi}_{11} + \boldsymbol{R}_{t-1}^{(p)\top}\boldsymbol{\psi}_{21} + \epsilon_{1t}\right)\right) = Aq, \\
\mathbb{E}\left(\boldsymbol{Z}_{t-1}^{(p)} M_t\right) &= \mathbb{E}\left(\boldsymbol{Z}_{t-1}^{(p)}\left(Z_t A + \boldsymbol{Z}_{t-1}^{(p)\top}\boldsymbol{\phi}_1 + \boldsymbol{M}_{t-1}^{(p)\top}\boldsymbol{\psi}_{11} + \boldsymbol{R}_{t-1}^{(p)\top}\boldsymbol{\psi}_{21} + \epsilon_{1t}\right)\right) \\
&= q\boldsymbol{\phi}_1 + Aq\boldsymbol{\psi}_{11} + (C + AB)q\boldsymbol{\psi}_{21}, \\
\mathbb{E}\left(\boldsymbol{M}_{t-1}^{(p)} M_t\right) &= \mathbb{E}\left(\boldsymbol{M}_{t-1}^{(p)}\left(Z_t A + \boldsymbol{Z}_{t-1}^{(p)\top}\boldsymbol{\phi}_1 + \boldsymbol{M}_{t-1}^{(p)\top}\boldsymbol{\psi}_{11} + \boldsymbol{R}_{t-1}^{(p)\top}\boldsymbol{\psi}_{21} + \epsilon_{1t}\right)\right) \\
&= Aq\boldsymbol{\phi}_1 + \boldsymbol{Q}_{\boldsymbol{M}\boldsymbol{M}}\boldsymbol{\psi}_{11} + \boldsymbol{Q}_{\boldsymbol{M}\boldsymbol{R}}\boldsymbol{\psi}_{21}, \\
\mathbb{E}\left(\boldsymbol{R}_{t-1}^{(p)} M_t\right) &= \mathbb{E}\left(\boldsymbol{R}_{t-1}^{(p)}\left(Z_t A + \boldsymbol{Z}_{t-1}^{(p)\top}\boldsymbol{\phi}_1 + \boldsymbol{M}_{t-1}^{(p)\top}\boldsymbol{\psi}_{11} + \boldsymbol{R}_{t-1}^{(p)\top}\boldsymbol{\psi}_{21} + \epsilon_{1t}\right)\right) \\
&= (C + AB)q\boldsymbol{\phi}_1 + \boldsymbol{Q}_{\boldsymbol{M}\boldsymbol{R}}^\top\boldsymbol{\psi}_{11} + \boldsymbol{Q}_{\boldsymbol{R}\boldsymbol{R}}\boldsymbol{\psi}_{21}.
\end{aligned}$$

For $\mathbb{E}[M_t^2]$, since

$$
\begin{aligned}
&\mathbb{E}(M_t^2)\\
&= \mathbb{E}(Z_t A + \boldsymbol{\phi}_1^\top \boldsymbol{Z}_{t-1}^{(p)} + \boldsymbol{\psi}_{11}^\top \boldsymbol{M}_{t-1}^{(p)} + \boldsymbol{\psi}_{21}^\top \boldsymbol{R}_{t-1}^{(p)} + \epsilon_{1t})(Z_t A + \boldsymbol{Z}_{t-1}^{(p)\top} \boldsymbol{\phi}_1 + \boldsymbol{M}_{t-1}^{(p)\top} \boldsymbol{\psi}_{11} + \boldsymbol{R}_{t-1}^{(p)\top} \boldsymbol{\psi}_{21} + \epsilon_{1t})\\
&= A^2 q + \sigma_1^2 + \boldsymbol{\phi}_1^\top \mathbb{E}\left(\boldsymbol{Z}_{t-1}^{(p)} \boldsymbol{Z}_{t-1}^{(p)\top}\right) \boldsymbol{\phi}_1 + \boldsymbol{\phi}_1^\top \mathbb{E}\left(\boldsymbol{Z}_{t-1}^{(p)} \boldsymbol{M}_{t-1}^{(p)\top}\right) \boldsymbol{\psi}_{11} + \boldsymbol{\phi}_1^\top \mathbb{E}\left(\boldsymbol{Z}_{t-1}^{(p)} \boldsymbol{R}_{t-1}^{(p)\top}\right) \boldsymbol{\psi}_{21}\\
&\quad + \boldsymbol{\psi}_{11}^\top \mathbb{E}\left(\boldsymbol{M}_{t-1}^{(p)} \boldsymbol{Z}_{t-1}^{(p)\top}\right) \boldsymbol{\phi}_1 + \boldsymbol{\psi}_{11}^\top \mathbb{E}\left(\boldsymbol{M}_{t-1}^{(p)} \boldsymbol{M}_{t-1}^{(p)\top}\right) \boldsymbol{\psi}_{11} + \boldsymbol{\psi}_{11}^\top \mathbb{E}\left(\boldsymbol{M}_{t-1}^{(p)} \boldsymbol{R}_{t-1}^{(p)\top}\right) \boldsymbol{\psi}_{21}\\
&\quad + \boldsymbol{\psi}_{21}^\top \mathbb{E}\left(\boldsymbol{R}_{t-1}^{(p)} \boldsymbol{Z}_{t-1}^{(p)\top}\right) \boldsymbol{\phi}_1 + \boldsymbol{\psi}_{21}^\top \mathbb{E}\left(\boldsymbol{R}_{t-1}^{(p)} \boldsymbol{M}_{t-1}^{(p)\top}\right) \boldsymbol{\psi}_{11} + \boldsymbol{\psi}_{21}^\top \mathbb{E}\left(\boldsymbol{R}_{t-1}^{(p)} \boldsymbol{R}_{t-1}^{(p)\top}\right) \boldsymbol{\psi}_{21},
\end{aligned}
$$

using the quantities derived above, we can calculate the Fisher information matrix $I(\boldsymbol{\theta})$ as well as the asymptotic covariance matrix. □

## A.5 Asymptotic property of $\widehat{AB}_p$

**Corollary A.2.** *Under the same condition of Theorem A.1, the asymptotic distribution of $\widehat{AB}_p$ is*

$$
\sqrt{T}\left(\widehat{AB}_p - AB\right) \xrightarrow{\mathcal{D}} \mathcal{N}\left(0, \boldsymbol{\theta}^\top \mathbf{J}_{1d} I^{-1}(\boldsymbol{\theta}) \mathbf{J}_{1d} \boldsymbol{\theta}\right), \tag{4}
$$

*where $d = 6p + 3$ is the dimension of $\boldsymbol{\theta}$,*

$$
\mathbf{J}_{1d} = \boldsymbol{e}_1 \boldsymbol{e}_d^\top + \boldsymbol{e}_d \boldsymbol{e}_1^\top
$$

*with $\boldsymbol{e}_1 = (1, 0, \ldots, 0)^\top$ and $\boldsymbol{e}_d = (0, \ldots, 0, 1)^\top$, thus the $(1, d)$ and $(d, 1)$ elements in $\mathbf{J}_{1d}$ are one and the rest are zero; and $I^{-1}(\boldsymbol{\theta})$ is the inverse Fisher information in Theorem A.4.*

*Proof.* Based on the definition of $\boldsymbol{\theta}$,

$$
A = \boldsymbol{e}_1^\top \boldsymbol{\theta}, \quad B = \boldsymbol{e}_d^\top \boldsymbol{\theta},
$$

then

$$AB \triangleq g\left(\boldsymbol{\theta}\right) = \left(\boldsymbol{e}_1^\top \boldsymbol{\theta}\right)\left(\boldsymbol{e}_d^\top \boldsymbol{\theta}\right) = \boldsymbol{e}_1^\top \left(\boldsymbol{\theta}\boldsymbol{\theta}^\top\right)\boldsymbol{e}_d. \tag{5}$$

The multivariate Delta method gives

$$\sqrt{T}\left(g(\hat{\boldsymbol{\theta}}) - g(\boldsymbol{\theta})\right) \xrightarrow{\mathcal{D}} \mathcal{N}\left(0, \nabla g(\boldsymbol{\theta})^\top I^{-1}(\boldsymbol{\theta}) \nabla g(\boldsymbol{\theta})\right), \tag{6}$$

where

$$\nabla g(\boldsymbol{\theta}) = \left(\boldsymbol{e}_1^\top \boldsymbol{e}_d + \boldsymbol{e}_d^\top \boldsymbol{e}_1\right)\boldsymbol{\theta} = \mathbf{J}_{1d}\boldsymbol{\theta}.$$

This proves the corollary. □

## A.6 Proof of Theorem 1

*Proof.* Plug the estimators in Theorem 2 into the conditional likelihood (10), we have

$$\begin{aligned} &\ell\left(\hat{\boldsymbol{\theta}}_1, \hat{\boldsymbol{\theta}}_2, \hat{B}, \hat{\sigma}_1, \hat{\sigma}_2\right) \\ =& -\frac{T-p}{2}\log\left(\frac{1}{(T-p)^2}\left(\boldsymbol{M}^\top(\mathbf{I} - \boldsymbol{P}_X)\boldsymbol{M}\right)\left(\boldsymbol{R}^\top(\mathbf{I} - \boldsymbol{P}_{MX} - \boldsymbol{P}_M)\boldsymbol{R}\right)\right) - (T-p), \end{aligned}$$

which is a constant function of $\delta$. □

## A.7 Proof of Theorem 3

*Proof.* In the single level model, for each participant, under assumption (A6), the estimators of the coefficients in the reparametrized models (14) and (15) are

$$
\begin{aligned}
\hat{\boldsymbol{\theta}}_{i_1} &= (\boldsymbol{X}_i^\top \boldsymbol{X}_i)^{-1}\boldsymbol{X}_i^\top \boldsymbol{M}_i,\\
\hat{\boldsymbol{\theta}}_{i_2} &= \left(\boldsymbol{X}_i^\top(\mathbf{I}-\boldsymbol{P}_{M_i})\boldsymbol{X}_i\right)^{-1}\boldsymbol{X}_i^\top(\mathbf{I}-\boldsymbol{P}_{M_i})\boldsymbol{R}_i + \frac{\delta}{\sqrt{1-\delta^2}}\sqrt{\frac{\boldsymbol{R}_i^\top(\mathbf{I}-\boldsymbol{P}_{MX_i}-\boldsymbol{P}_{M_i})\boldsymbol{R}_i}{\boldsymbol{M}_i^\top(\mathbf{I}-\boldsymbol{P}_{X_i})\boldsymbol{M}_i}}(\boldsymbol{X}_i^\top \boldsymbol{X}_i)^{-1}\boldsymbol{X}_i^\top \boldsymbol{M}_i\\
&\triangleq \tilde{\boldsymbol{\theta}}_{i_2} + \tau\boldsymbol{\xi}_{i_2},\\
\hat{B}_i &= (\boldsymbol{M}_i^\top \boldsymbol{M}_i)^{-1}\boldsymbol{M}_i\left(\mathbf{I}-\boldsymbol{X}_i(\boldsymbol{X}_i^\top(\mathbf{I}-\boldsymbol{P}_{M_i})\boldsymbol{X}_i)^{-1}\boldsymbol{X}_i^\top(\mathbf{I}-\boldsymbol{P}_{M_i})\right)\boldsymbol{R}_i\\
&\quad - \frac{\delta}{\sqrt{1-\delta^2}}\sqrt{\frac{\boldsymbol{R}_i^\top(\mathbf{I}-\boldsymbol{P}_{MX_i}-\boldsymbol{P}_{M_i})\boldsymbol{R}_i}{\boldsymbol{M}_i^\top(\mathbf{I}-\boldsymbol{P}_{X_i})\boldsymbol{M}_i}}\\
&\triangleq \tilde{B}_i - \tau b_i,
\end{aligned}
$$

where

$$
\tau = \frac{\delta}{\sqrt{1-\delta^2}}, \quad b_i = \sqrt{\frac{\boldsymbol{R}_i^\top(\mathbf{I}-\boldsymbol{P}_{MX_i}-\boldsymbol{P}_{M_i})\boldsymbol{R}_i}{\boldsymbol{M}_i^\top(\mathbf{I}-\boldsymbol{P}_{X_i})\boldsymbol{M}_i}}, \quad \boldsymbol{\xi}_{i_2} = b_i\hat{\boldsymbol{\theta}}_{i_1}.
$$

$C_i$ is the first element in $\boldsymbol{\theta}_{i_2}$, thus

$$
\hat{C}_i = \boldsymbol{e}_1^\top\hat{\boldsymbol{\theta}}_{i_2} = \boldsymbol{e}_1^\top\tilde{\boldsymbol{\theta}}_{i_2} + \tau\boldsymbol{e}_1^\top\boldsymbol{\xi}_{i_2} \triangleq \tilde{C}_i + \tau c_i,
$$

where $\boldsymbol{e}_1$ is a $(3p+1)\times 1$ vector with first element one and the rest zero. Following the notations in the proof of Theorem 3 of Zhao and Luo (2014), by maximizing the profile likelihood of $\tau$ in model (13), we have

(1) if $\boldsymbol{\Lambda}$ is known,

$$
\hat{\tau} = \frac{\boldsymbol{\Delta}_{\tilde{B},b}/\lambda_\beta^2 - \boldsymbol{\Delta}_{\tilde{C},c}/\lambda_\gamma^2}{\boldsymbol{\Delta}_b/\lambda_\beta^2 + \boldsymbol{\Delta}_c/\lambda_\gamma^2};
$$

(2) if $\boldsymbol{\Lambda}$ is unknown, the estimator of $\tau$ should satisfy the following cubic function

$$2\tau^3\boldsymbol{\Delta}_b\boldsymbol{\Delta}_c+\tau^2(3\boldsymbol{\Delta}_b\boldsymbol{\Delta}_{\tilde{C},c}-3\boldsymbol{\Delta}_{\tilde{B},b}\boldsymbol{\Delta}_c)+\tau(\boldsymbol{\Delta}_{\tilde{B}}\boldsymbol{\Delta}_c+\boldsymbol{\Delta}_b\boldsymbol{\Delta}_{\tilde{C}}-4\boldsymbol{\Delta}_{\tilde{B},b}\boldsymbol{\Delta}_{\tilde{C},c})+(\boldsymbol{\Delta}_{\tilde{B}}\boldsymbol{\Delta}_{\tilde{C},c}-\boldsymbol{\Delta}_{\tilde{B},b}\boldsymbol{\Delta}_{\tilde{C}})=0.$$

Implementing the same strategy as in Zhao and Luo (2014), for the single level model, under the stationary condition,

$$b_i=\sqrt{\frac{\boldsymbol{R}_i^\top(\mathbf{I}-\boldsymbol{P}_{MX_i}-\boldsymbol{P}_{M_i})\boldsymbol{R}_i}{\boldsymbol{M}_i^\top(\mathbf{I}-\boldsymbol{P}_{X_i})\boldsymbol{M}_i}},\text{ and } b_i\mid\boldsymbol{\theta}_{i_1},\boldsymbol{\theta}_{i_2},B_i,\boldsymbol{\Sigma}_i=\sqrt{1-\delta^2}\frac{\sigma_{i_2}}{\sigma_{i_1}}+\mathcal{O}_p\left(\frac{1}{\sqrt{T_i}}\right),$$

where $T_i$ is the number of time points of participant $i$, $i=1,\ldots,N$. The rest of the proof follows (for details, see the proof of Theorem 3 in Zhao and Luo (2014)). □

## A.8 Proof of Corollary 1

Given the results in Theorem A.1 and Theorem 3, the conclusion for parameters in the first-level model is straightforward following the Slutsky's theorem. For the population level parameter, conclusion for estimator of $A$ is trivial as $\hat{A}_i$'s are in independent of $\delta$. For $B$ and $C$, as $\hat{B}_i$ and $\hat{C}_i$ are functions of $\hat{\delta}$, $\hat{B}_i$ and $\hat{B}_{i'}$ for $i\neq i'$ are dependent, same for $\hat{C}_i$'s. From Theorem 3, $\hat{\delta}-\delta=\mathcal{O}_p(1/\sqrt{NT})$, and we have $\mathrm{Cov}(\hat{B}_i,\hat{B}_{i'})=\mathrm{Cov}(\tilde{B}_i-\hat{\delta}\hat{\sigma}_{i_2}/\hat{\sigma}_{i_1},\tilde{B}_{i'}-\hat{\delta}\hat{\sigma}_{i'_2}/\hat{\sigma}_{i'_1})\to 0$ as $N\to\infty,T\to\infty$, where $\tilde{B}_i$ is the estimator when assuming $\delta=0$ and thus independent across subjects. Based on the results in Kozlov et al. (2004), the estimator of population-level parameter $B$ is consistent. Proof for $C$ can be derived analogously.

## A.9 Proof of Theorem 4

*Proof.* Using the same matrix notation as in Section 2.3 for participant $i$, the transformed single level model is written as

$$
\begin{aligned}
M_{i_t} &= \boldsymbol{X}_{i_t}^\top \boldsymbol{\theta}_{i_1} + \epsilon_{i_1 t}, \\
R_{i_t} &= M_{i_t} B_i + \boldsymbol{X}_{i_t}^\top \boldsymbol{\theta}_{i_2} + \epsilon_{i_2 t}.
\end{aligned} \tag{7}
$$

Let

$$
\boldsymbol{R}_i = \begin{pmatrix} R_{i_{p+1}} \\ \vdots \\ R_{i_{T_i}} \end{pmatrix}, \quad \boldsymbol{M}_i = \begin{pmatrix} M_{i_{p+1}} \\ \vdots \\ M_{i_{T_i}} \end{pmatrix}, \quad \boldsymbol{X}_i = \begin{pmatrix} \boldsymbol{X}_{i_{p+1}}^\top \\ \vdots \\ \boldsymbol{X}_{i_{T_i}}^\top \end{pmatrix}, \quad \boldsymbol{\theta}_i = \begin{pmatrix} \boldsymbol{\theta}_{i_1} \\ \boldsymbol{\theta}_{i_2} \\ B_i \end{pmatrix},
$$

then

$$
\begin{aligned}
h(\boldsymbol{\Upsilon}) &= \sum_{i=1}^{N} \sum_{t=1}^{T} \log \mathbb{P}\left(R_{i_t}, M_{i_t} \mid Z_{i_t}, \boldsymbol{Z}_{i_{t-1}}^{(p)}, \boldsymbol{M}_{i_{t-1}}^{(p)}, \boldsymbol{R}_{i_{t-1}}^{(p)}, \boldsymbol{\theta}_{i_1}, \boldsymbol{\theta}_{i_2}, B_i, \delta, \sigma_{i_1}, \sigma_{i_2}\right) + \sum_{i=1}^{N} \log \mathbb{P}\left(b_i \mid b, \boldsymbol{\Lambda}\right) \\
&= -\frac{1}{2} \sum_{i=1}^{N} \Big[ (T_i - p) \log \sigma_{i_1}^2 \sigma_{i_2}^2 (1 - \delta^2) + \frac{1}{\sigma_{i_1}^2} (\boldsymbol{M}_i - \boldsymbol{X}_i \boldsymbol{J}_1 \boldsymbol{\theta}_i)^\top (\boldsymbol{M}_i - \boldsymbol{X}_i \boldsymbol{J}_1 \boldsymbol{\theta}_i) \\
&\quad + \frac{1}{\sigma_{i_2}^2 (1 - \delta^2)} \big((\boldsymbol{R}_i - \boldsymbol{M}_i \boldsymbol{J}_3 \boldsymbol{\theta}_i - \boldsymbol{X}_i \boldsymbol{J}_2 \boldsymbol{\theta}_i) - \kappa_i (\boldsymbol{M}_i - \boldsymbol{X}_i \boldsymbol{J}_1 \boldsymbol{\theta}_i)\big)^\top \\
&\quad \big((\boldsymbol{R}_i - \boldsymbol{M}_i \boldsymbol{J}_3 \boldsymbol{\theta}_i - \boldsymbol{X}_i \boldsymbol{J}_2 \boldsymbol{\theta}_i) - \kappa_i (\boldsymbol{M}_i - \boldsymbol{X}_i \boldsymbol{J}_1 \boldsymbol{\theta}_i)\big) \Big] \\
&\quad - \frac{1}{2} \sum_{i=1}^{N} \Big[ \log |\boldsymbol{\Lambda}| + (\boldsymbol{J} \boldsymbol{\theta}_i - b)^\top \boldsymbol{\Lambda}^{-1} (\boldsymbol{J} \boldsymbol{\theta}_i - b) \Big],
\end{aligned}
$$

where $T_i$ is the number of time points of participant $i$; $\kappa_i = \delta \sigma_{i_2} / \sigma_{i_1}$; and

$$
\boldsymbol{J}_1 = \begin{pmatrix} \mathbf{I}_{3p+1} & \mathbf{0}_{3p+1} & \mathbf{0} \end{pmatrix}_{(3p+1)\times(6p+3)}, \; \boldsymbol{J}_2 = \begin{pmatrix} \mathbf{0}_{3p+1} & \mathbf{I}_{3p+1} & \mathbf{0} \end{pmatrix}_{(3p+1)\times(6p+3)},
$$

$$
\boldsymbol{J}_3 = \begin{pmatrix} \mathbf{0}^\top & \mathbf{0}^\top & \mathbf{0} \end{pmatrix}_{1\times(6p+3)}, \ \boldsymbol{J} = \begin{pmatrix} 1 & 0 & \cdots & 0 & 0 & 0 & \cdots & 0 & 0 \\ 0 & 0 & \cdots & 0 & 0 & 0 & \cdots & 0 & 1 \\ 0 & 0 & \cdots & 0 & 1 & 0 & \cdots & 0 & 0 \end{pmatrix}_{3\times(6p+3)}.
$$

We first prove the conditional convexity of $h$. For $\left(\sigma_{i_1}^{-1}, \sigma_{i_2}^{-1}\right)$,

$$
\begin{aligned}
\frac{\partial^2(-h)}{\partial \sigma_{i_1}^{-2}} &= \frac{T_i - p}{\sigma_{i_1}^{-2}} + \frac{1}{1-\delta^2} \boldsymbol{e}_{i_1}^\top \boldsymbol{e}_{i_1}, \\
\frac{\partial^2(-h)}{\partial \sigma_{i_2}^{-2}} &= \frac{T_i - p}{\sigma_{i_2}^{-2}} + \frac{1}{1-\delta^2} \boldsymbol{e}_{i_2}^\top \boldsymbol{e}_{i_2}, \\
\frac{\partial^2(-h)}{\partial \sigma_{i_1}^{-1} \sigma_{i_2}^{-1}} &= -\frac{\delta}{1-\delta^2} \boldsymbol{e}_{i_1}^\top \boldsymbol{e}_{i_2},
\end{aligned}
$$

where

$$
\begin{aligned}
\boldsymbol{e}_{i_1} &= \boldsymbol{M}_i - \boldsymbol{X}_i \boldsymbol{J}_1 \boldsymbol{\theta}_i, \\
\boldsymbol{e}_{i_2} &= \boldsymbol{R}_i - \boldsymbol{M}_i \boldsymbol{J}_3 \boldsymbol{\theta}_i - \boldsymbol{X}_i \boldsymbol{J}_2 \boldsymbol{\theta}_i.
\end{aligned} \tag{8}
$$

Let

$$
\boldsymbol{\Delta} = \begin{pmatrix} 1 & \delta \\ \delta & 1 \end{pmatrix}, \quad \boldsymbol{S}_i = \begin{pmatrix} \boldsymbol{e}_{i_1}^\top \boldsymbol{e}_{i_1} & \boldsymbol{e}_{i_1}^\top \boldsymbol{e}_{i_2} \\ \boldsymbol{e}_{i_2}^\top \boldsymbol{e}_{i_1} & \boldsymbol{e}_{i_2}^\top \boldsymbol{e}_{i_2} \end{pmatrix},
$$

then the Hessian matrix is

$$
\begin{aligned}
&\begin{pmatrix} (T_i - p)/\sigma_{i_1}^{-2} + \boldsymbol{\Delta}^{-1}(1,1)\boldsymbol{S}_i(1,1) & \boldsymbol{\Delta}^{-1}(1,2)\boldsymbol{S}_i(1,2) \\ \boldsymbol{\Delta}^{-1}(2,1)\boldsymbol{S}_i(2,1) & (T_i - p)/\sigma_{i_2}^{-2} + \boldsymbol{\Delta}^{-1}(2,2)\boldsymbol{S}_i(2,2) \end{pmatrix} \\
= &\begin{pmatrix} (T_i - p)/\sigma_{i_1}^{-2} & 0 \\ 0 & (T_i - p)/\sigma_{i_2}^{-2} \end{pmatrix} + \boldsymbol{\Delta}^{-1} \circ \boldsymbol{S}_i,
\end{aligned} \tag{9}
$$

where $\boldsymbol{\Delta}^{-1} \circ \boldsymbol{S}_i$ is the Hadamard product of $\boldsymbol{\Delta}^{-1}$ and $S_i$, which is positive semidefinite. Therefore, the negative full likelihood $(-h)$ is convex in $\left(\sigma_{i_1}^{-1}, \sigma_{i_2}^{-1}\right)$ conditional on the rest parameters.

For $\boldsymbol{\theta}_i$ and $b$, the second-order partial derivatives are

$$\frac{\partial^2(-h)}{\partial\boldsymbol{\theta}_i\partial\boldsymbol{\theta}_i^\top} = \frac{1}{\sigma_{i_1}^2}\boldsymbol{J}_1^\top\boldsymbol{X}_i^\top\boldsymbol{X}_i\boldsymbol{J}_1 + \frac{1}{\sigma_{i_2}^2(1-\delta^2)}\left(\boldsymbol{M}_i\boldsymbol{J}_3 + \boldsymbol{X}_i\boldsymbol{J}_2 - \kappa_i\boldsymbol{X}_i\boldsymbol{J}_1\right)^\top\left(\boldsymbol{M}_i\boldsymbol{J}_3 + \boldsymbol{X}_i\boldsymbol{J}_2 - \kappa_i\boldsymbol{X}_i\boldsymbol{J}_1\right) + \boldsymbol{J}^\top\boldsymbol{\Lambda}^{-1}\boldsymbol{J},$$
$$\frac{\partial^2(-h)}{\partial b\partial b^\top} = N\boldsymbol{\Lambda}^{-1},$$

both of which are positive semidefinite.

For $\boldsymbol{\Lambda}^{-1}$, the second-order partial derivative can be calculated as

$$\partial\left(\frac{\partial(-h)}{\partial\boldsymbol{\Lambda}^{-1}}\right) = \frac{N}{2}\boldsymbol{\Lambda}\left(\partial\boldsymbol{\Lambda}^{-1}\right)\boldsymbol{\Lambda}.$$

This finishes the proof of the conditional convexity of the negative likelihood function (16) when $\delta$ is given.

Next, we show that given $\delta$, the optimization for $h$ is separable and derive the optimizers in explicit forms. After taking the first-order partial derivatives, the optimizers are

$$\hat{\sigma}_{i_1}^2 = \frac{1}{(T_i-p)(1-\delta^2)}\left[\boldsymbol{S}_i(1,1) - \delta\boldsymbol{S}_i(1,2)\sqrt{\frac{\boldsymbol{S}_i(1,1)}{\boldsymbol{S}_i(2,2)}}\right],$$
$$\hat{\sigma}_{i_2}^2 = \frac{1}{(T_i-p)(1-\delta^2)}\left[\boldsymbol{S}_i(2,2) - \delta\boldsymbol{S}_i(2,1)\sqrt{\frac{\boldsymbol{S}_i(2,2)}{\boldsymbol{S}_i(1,1)}}\right],$$
$$\hat{\boldsymbol{\theta}}_i = \left[\frac{1}{\sigma_{i_1}^2}\boldsymbol{J}_1^\top\boldsymbol{X}_i^\top\boldsymbol{X}_i\boldsymbol{J}_1 + \frac{1}{\sigma_{i_2}^2(1-\delta^2)}(\boldsymbol{M}_i\boldsymbol{J}_3 + \boldsymbol{X}_i\boldsymbol{J}_2 - \kappa_i\boldsymbol{X}_i\boldsymbol{J}_1)^\top(\boldsymbol{M}_i\boldsymbol{J}_3 + \boldsymbol{X}_i\boldsymbol{J}_2 - \kappa_i\boldsymbol{X}_i\boldsymbol{J}_1) + \boldsymbol{J}^\top\boldsymbol{\Lambda}^{-1}\boldsymbol{J}\right]^{-1}$$
$$\left[\frac{1}{\sigma_{i_1}^2}\boldsymbol{J}_1^\top\boldsymbol{X}_i^\top\boldsymbol{M}_i + \frac{1}{\sigma_{i_2}^2(1-\delta^2)}(\boldsymbol{M}_i\boldsymbol{J}_3 + \boldsymbol{X}_i\boldsymbol{J}_2 - \kappa_i\boldsymbol{X}_i\boldsymbol{J}_1)^\top(\boldsymbol{R}_i - \kappa_i\boldsymbol{M}_i) + \boldsymbol{J}^\top\boldsymbol{\Lambda}^{-1}b\right],$$
$$\hat{b} = \frac{1}{N}\sum_{i=1}^{N}\boldsymbol{J}\boldsymbol{\theta}_i,$$
$$\hat{\boldsymbol{\Lambda}} = \frac{1}{N}\sum_{i=1}^{N}(\boldsymbol{J}\boldsymbol{\theta}_i - b)(\boldsymbol{J}\boldsymbol{\theta}_i - b)^\top.$$

Since $\mathcal{S}$ is a convex set, if the updates of the variance components are interior points of $\mathcal{S}$, the updating formula above will be applied; otherwise the solutions will be projected onto the set $\mathcal{S}$. □

# B Additional Simulation Results

## B.1 Estimate of $B$ and $C$ in the setting of Figure 5 with varying $\delta$. (Figure B.1)

The estimates of $B$ and $C$ with varying $\delta$ are shown in Figure B.1. GMA-h yields lower bias

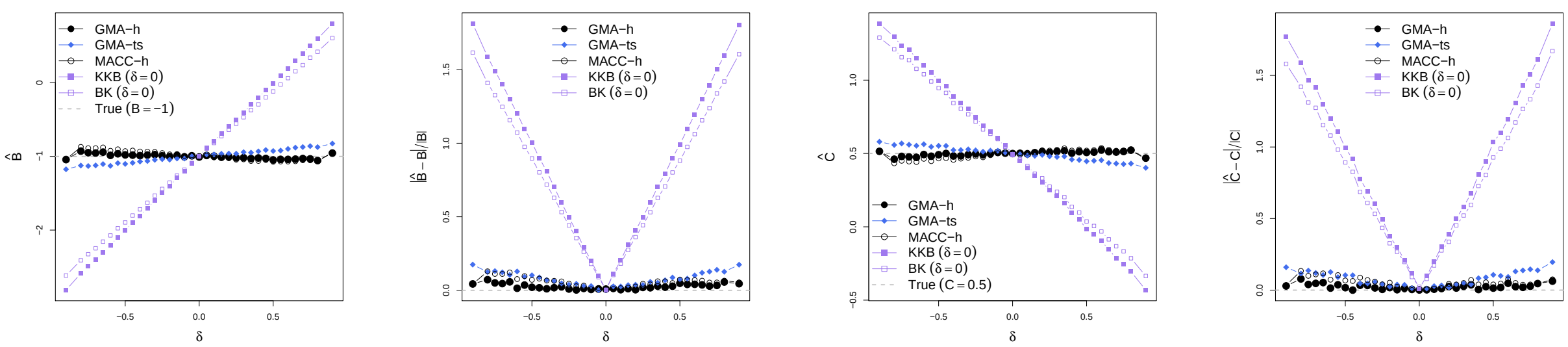

(a) $\hat{B}$. (b) $|\hat{B}-B|/|B|$. (c) $\hat{C}$. (d) $|\hat{C}-C|/|C|$.

Figure B.1: Point estimate $B$ and $C$ and the relative bias under different true $\delta$ values.

than CMA-h and GMA-ts. The bias of KKB and BK increase as $|\delta|$ increases.

## B.2 Estimate of $B$ and $C$ in the setting of Figure 6. (Figure B.2)

Figure B.2 shows the estimates and the corresponding MSE in estimating $B$ and $C$ from GMA-h and GMA-ts as $N$ and $T_i$ increase. From the figure, the estimates from both methods

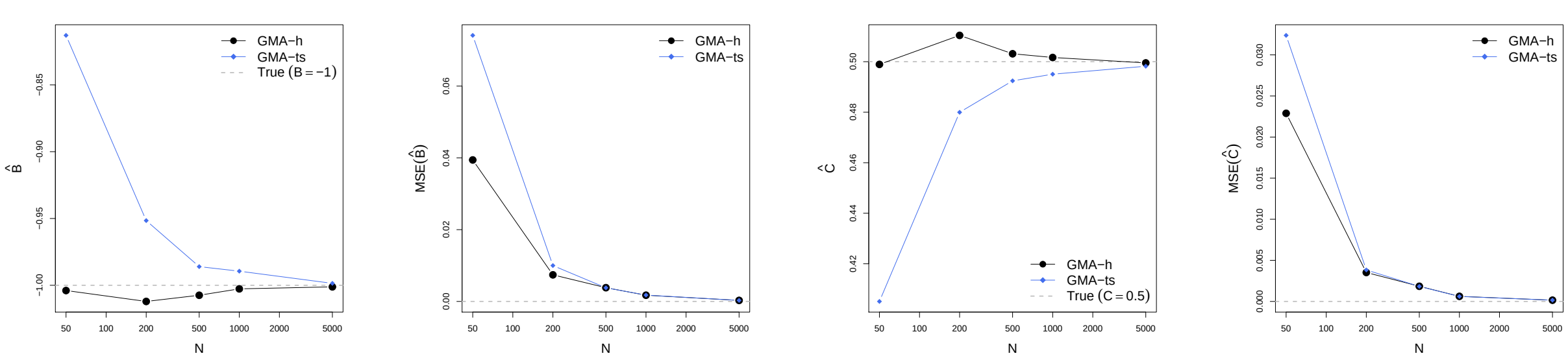

(a) $\hat{B}$. (b) MSE of $\hat{B}$. (c) $\hat{C}$. (d) MSE of $\hat{C}$.

Figure B.2: Average point estimates and MSEs of $\hat{B}$ and $\hat{C}$ from GMA-h and GMA-ts.

converge to the true value as sample sizes increase. GMA-h has lower bias than the GMA-ts does.

# C Additional results of the fMRI study

## C.1 Estimate of $\Omega$ using GMA-h and Granger causality (Harrison et al., 2003)

Table C.1 presents the estimate, as well as the corresponding 95% confidence interval, of the transition matrix $\mathbf{\Omega}$ with $p = 2$ using our GMA-h and the Granger causality approaches. From the table, the estimate of off-diagonal elements from the two approaches are very close.

Table C.1: The estimates of the transition matrix $\mathbf{\Omega}$ and the corresponding 95% confidence intervals (CIs) using the GMA-h and the Granger causality methods, from 200 participant-level bootstrap samples.

| | GMA-h | | Granger causality | |
|---|---|---|---|---|
| | Estimate | 95% CI | Estimate | 95% CI |
| $\omega_{11_1}$ | 0.436 | (0.401, 0.426) | 0.359 | (0.316, 0.404) |
| $\omega_{12_1}$ | 0.133 | (0.104, 0.161) | 0.135 | (0.097, 0.167) |
| $\omega_{21_1}$ | 0.100 | (0.072, 0.132) | 0.100 | (0.074, 0.130) |
| $\omega_{22_1}$ | 0.357 | (0.321, 0.395) | 0.460 | (0.427, 0.493) |
| $\omega_{11_2}$ | -0.095 | (-0.119, -0.071) | -0.033 | (-0.058, -0.005) |
| $\omega_{12_2}$ | -0.019 | (-0.034, -0.001) | -0.029 | (-0.051, -0.006) |
| $\omega_{21_2}$ | -0.076 | (-0.096, -0.053) | -0.074 | (-0.092, -0.051) |
| $\omega_{22_2}$ | -0.025 | (-0.046, -0.003) | -0.089 | (-0.114, -0.063) |

However, we observe significant difference in the estimate of diagonal elements.

## C.2 Estimate of $\Omega$ under a MAR(3) model for the errors.

Figure C.1 shows the estimates of $\mathbf{\Omega}_1$, $\mathbf{\Omega}_2$ and $\mathbf{\Omega}_3$ under MAR(3) model for data without motion correction using the GMA-h method. From the figure, the estimates of $\mathbf{\Omega}_1$ and $\mathbf{\Omega}_2$

are very close to the estimates under MAR(2) model, and the estimates of the elements in $\mathbf{\Omega}_3$ are close to zero. This suggests that for our fMRI dataset, MAR(1) is sufficient for the errors in model (11).

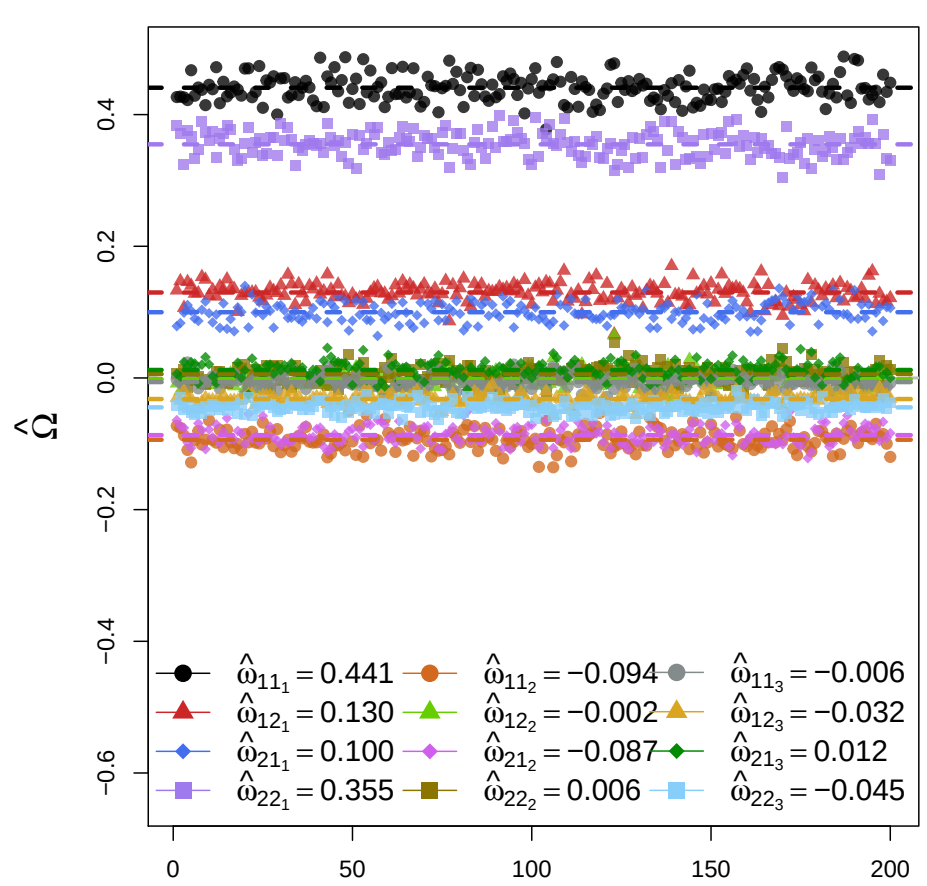

Figure C.1: The estimates of the transition matrix $\boldsymbol{\Omega}$ for data without motion correction from 500 subject-level bootstrap samples under MAR(3) model using the GMA-h method.